\documentclass[12pt]{article}
\usepackage{graphicx} 
\usepackage{amsmath} 
\usepackage{xcolor}
\usepackage{subcaption}  
\usepackage{float}   
\usepackage{natbib}
\usepackage{bibunits}
\usepackage{booktabs}
\usepackage{caption}
\usepackage{bm}
\usepackage{placeins}
\usepackage{amsfonts}
\usepackage[USenglish]{babel}
\usepackage{mathptmx}
\usepackage[indent=0pt, skip=10pt]{parskip}
\usepackage[letterpaper, margin=1in]{geometry}

\usepackage[colorlinks=true, linkcolor=black, citecolor=black, urlcolor=black]{hyperref}

\usepackage[justification=centering]{caption}
\usepackage[justification=centering]{subcaption}
\usepackage{ragged2e}
\RaggedRight
\usepackage{lineno}
\usepackage{setspace}
\usepackage{changepage}

\renewenvironment{abstract}{%
    \small
    \begin{adjustwidth}{0cm}{0cm}%
    \begin{center}%
    \bfseries \abstractname\vspace{-.5em}\vspace{0pt}%
    \end{center}%
}{%
    \end{adjustwidth}
}

\usepackage{titlesec}

\titlespacing*{\section}
{0pt}      
{0.2em}    
{0.6em}    

\titlespacing*{\subsection}
{0pt}      
{0.3em}    
{0.6em}    

\title{\bfseries
Inferring resource selection and utilization distributions from irregular and error-prone animal tracking data}

\author{
Fanny Dupont$^{1,*}$,
Brett T.\ McClintock$^{2}$,
Jan-Ole Fischer$^{3}$,
Marianne Marcoux$^{4}$,\\
Nigel Hussey$^{5}$,
Marie Auger-Méthé$^{1,6}$
}

\date{} 

\begin{document}
\defaultbibliographystyle{chicago}  
\defaultbibliography{bibli}        
\begin{titlepage}
\pagenumbering{arabic}
\thispagestyle{plain}
\begin{spacing}{1}

\vspace{1.5em}

\begin{center}
{\LARGE\bfseries
Inferring resource selection and utilization distributions from irregular and error-prone animal tracking data\par}
\end{center}
\vspace{1em}
\begin{center}
{\large
Fanny Dupont$^{1*}$,
Brett T.\ McClintock$^{2,3}$,
Jan-Ole Fischer$^{4}$,
Marianne Marcoux$^{5}$,\\
Nigel E. Hussey$^{6}$,
Marie Auger-Méthé$^{1,7}$\par}
\end{center}
\vspace{0.75em}

\flushleft
{\small
$^{1}$Department of Statistics, University of British Columbia, Vancouver, British Columbia, Canada\\
$^{2}$Marine Mammal Laboratory, NOAA-NMFS Alaska Fisheries Science Center, Seattle, WA, USA\\
$^{3}$School of Environmental and Forest Sciences, University of Washington, Seattle, WA, USA\\
$^{4}$Department of Business Administration and Economics, Bielefeld University, Bielefeld, North Rhine-Westphalia, Germany\\
$^{5}$Freshwater Institute, Fisheries and Oceans Canada, Winnipeg, Manitoba, Canada\\
$^{6}$Department of Integrative Biology, University of Windsor, Windsor, Ontario, Canada\\
$^{7}$Institute for the Oceans and Fisheries, University of British Columbia, Vancouver, British Columbia, Canada\\[0.5em]
$^{*}$Corresponding author: \texttt{fanny.dupont@stat.ubc.ca}\par}

\vspace{1.25em}

\noindent\textbf{Open Research Statement:}  
An R package for simulating and fitting the Langevin SSM is available on GitHub at \url{https://github.com/bmcclintock/langevinSSM} and all code and data to reproduce the simulation study and data analysis is available publicly at \url{https://github.com/Fanny-Dupont/Langevin_SSM}.

\vspace{0.75em}

\noindent\textit{Key words/phrases:
animal movement; 
Argos;
Fastloc GPS;
habitat selection;
Langevin diffusion;
measurement error; 
step selection analysis
}

\end{spacing}
\end{titlepage}
\newpage
\begin{abstract}
Habitat selection and space use are fundamental to understanding animal distribution and informing conservation strategies. Traditional methods for quantifying habitat preferences and deriving utilization distributions from telemetry-derived location data assume regular sampling and negligible measurement error. These assumptions are routinely violated, which leads many practitioners to regularize and filter data before fitting models. These two-step procedures fail to propagate uncertainty from the filtering stage into subsequent analyses and can yield biased estimates. Langevin diffusion models can elegantly represent movement as a stochastic process driven by habitat-selection dynamics in continuous time, naturally accommodating irregular sampling. However, a state-space formulation that incorporates measurement error by treating true locations as latent variables remains challenging because habitat data depending on these locations are no longer observed. We facilitate the use of such state-space formulation by using the Laplace approximation to simultaneously integrate over the true locations and account for the habitat covariates encountered along those latent paths, yielding a single-stage framework that can be efficiently implemented with Template Model Builder (TMB). By doing so, we provide the first TMB implementation capable of handling covariates that depend on latent variables, allowing inference via fast and efficient maximum likelihood estimation. Simulation studies demonstrate that our approach outperforms the two-step method, recovering the sign and magnitude of habitat-selection parameters and yielding accurate utilization distribution and trajectory reconstructions even under the substantial measurement error and missing data levels common in marine studies. Our narwhal (\textit{Monodon monoceros}) case study demonstrates that the two-step method masks the habitat selection coefficient substantially towards zero, while our unified approach recovers a much stronger selection signal. By providing a solution for both measurement error and temporal irregularity when inferring habitat selection and utilization distributions, our approach addresses several long-standing challenges and offers a computationally efficient, flexible framework for movement ecologists that is applicable across a wide range of taxa and environments.

\end{abstract}
\begin{bibunit}
\section{Introduction}
\vspace{-0.3cm}
Animals do not move uniformly across their habitat, and mapping the probability of species presence across a landscape is fundamental in ecology, and for informing conservation and ecosystem management (\citealp{northrup_conceptual_2022,matthiopoulos2023species}). Habitat selection analyses aim to uncover these dynamics by quantifying which environmental features animals select relative to what is available to them, thereby indicating relative preference (or avoidance) for particular habitats (\citealp{manly2002resource, matthiopoulos2023species}). This idea underlies the concept of the utilization distribution (UD; \citealp{anderson_home_1982}), which provides a spatial map of relative intensity of space use, showing where the animal is most likely to be. The UD thus helps identifying home ranges (\citealp{sprogis_home_2016}), key habitats (\citealp{hebblewhite2009trade}), and zones of potential overlap with human activities (\citealp{halliday_potential_2021}). Such insights are vital for guiding conservation and management efforts (\citealp{mcwhinnie_vessel_2018}). Advances in tracking technology now provide the extensive, fine-scale movement data needed to perform these analyses across species and environments (\citealp{nathan2022big}).

Many methods have been developed to estimate habitat selection and UDs from telemetry data. Standard approaches fit a resource selection function (RSF) to quantify habitat selection (\citealp{boyce1999relating}; \citealp{manly2002resource}; \citealp{johnson_estimating_2013}; \citealp{hooten_animal_2017}; \citealp{matthiopoulos2023species}), and the UD is then derived as the normalized product of the RSF and an availability distribution, which describes the distribution of habitat covariates over locations accessible to the animal 
(\citealp{hooten_animal_2017, matthiopoulos2023species}). Although multiple methods exist for fitting RSFs to telemetry data, the most common are use–available logistic regression (\citealp{freitas2016temperature}) and point process models (\citealp{warton_poisson_2010}; \citealp{hooten_animal_2017}). Despite their widespread use, RSFs rely on the assumption of spatio-temporal independence among locations, which is increasingly violated by modern high-resolution movement data (\citealp{nathan2022big}). Statistical methods have therefore been developed to model how animals respond to environmental conditions while accounting for the spatio-temporal structure of movement data.

Among the several approaches proposed to account for autocorrelation in RSFs (e.g., data thinning, \citealp{northrup_practical_2013}; weighted likelihood, \citealp{alston_mitigating_2023}; and variance inflation, \citealp{nielsen2002modeling}), step-selection functions (SSFs) have emerged as a prominent solution (\citealp{thurfjell_applications_2014, avgar_integrated_2016, fieberg2021guide}). In SSFs, habitat availability is constrained by the animal’s movement, such that selection is evaluated only among locations that are reachable given the preceding position and movement characteristics. Step-selection functions therefore describe habitat selection at a local scale, and the parameter estimates are inherently tied to the temporal resolution of the sampled observations (\citealp{schlagel_robustness_2016}). While some ecological questions are focused on such local, short-term selection (e.g., how immediate predator presence affects habitat selection, \citealp{matthews2020killer}), this dependence limits the characterization of the species' broad scale space use generally captured by a  UD (\citealp{michelot_linking_2019}). Although step-scale selection can be translated into long-term space use, both theoretical and empirical work have shown that SSFs 
do not recover the UD implied by an RSF (\citealp{signer2017estimating}; \citealp{michelot_linking_2019}). 
Moreover, since parameter estimates depend on the sampling interval, SSF-derived inferences are not comparable across datasets with different temporal resolutions, nor easily related to broader-scale species distribution models.

Several approaches have been proposed to upscale inference from SSFs to the UD (\citealp{potts_how_2023}), though simulation-based procedures are most common (\citealp{signer2024simulating}). This reliance on simulation makes it difficult to obtain a simple parametric UD and complicates uncertainty quantification, which is typically ignored (\citealp{michelot_multiscale_2024}). Continuous-time stochastic processes offer an elegant solution, naturally accommodating irregularly sampled data while yielding a global UD from local movement decisions via their limiting distribution \citep[e.g.,][]{calabrese2016ctmm,scharf_imputation_2017, wilson2018estimating, michelot_linking_2019}. Among them, the habitat-driven Langevin diffusion \citep{michelot_langevin_2019,michelot_multiscale_2024} is particularly promising for habitat selection and UD inference across temporal scales. At the fine scale, the Langevin stochastic differential equation (SDE) models continuous movement, with a drift driven by local gradients in the habitat potential surface. This surface is defined by a standard RSF, ensuring that the stationary distribution of the location process (i.e., the UD) is strictly proportional to the RSF. The broad-scale UD thus emerges from continuous, fine-scale responses to these gradients over time. Growing interest in the Langevin model has since led to several extensions (\citealp{mcclintock_multistate_2024}; \citealp{delporte_spatial_2025}).


Standard habitat selection models, including the Langevin diffusion, do not account for location measurement error, which is pervasive in telemetry studies, particularly in marine systems relying on tracking technologies such as Advanced Research and Global Observation Satellite (Argos; \citealp{argos2017}) and Fastloc GPS (\citealp{costa_accuracy_2010}; \citealp{wildlifecomputers_fastloc}). Inferring habitat selection from noisy locations can result in bias and, ultimately, misleading conclusions. To address this, state-space models (SSMs) are routinely used as a preliminary filtering step. Implemented via \texttt{R} packages such as \texttt{aniMotum} (\citealp{jonsen_continuous-time_2020}), \texttt{ctmm} (\citealp{calabrese2016ctmm}) and \texttt{crawl} (\citealp{johnson2018crawl}), SSMs incorporate measurement error and predict the true movement track, typically modeled as a random walk. This two-step pipeline, correcting for error before performing ecological inference, has become standard in habitat selection analyses of error-prone data \citep[e.g.,][]{matthews2020killer}. However, most users do not propagate the uncertainty from the filtering step into subsequent analyses, potentially introducing bias, the extent of which has not been thoroughly investigated (\citealp{michelot_langevin_2019}). Such bias may arise, for example, when the movement model used for filtering is overly simplistic.

Building on the Langevin diffusion framework, we explicitly express the stationary distribution as an RSF and incorporate measurement error via an SSM. We propose a highly efficient likelihood-based fitting method implemented within Template Model Builder (TMB; \citealp{kristensen_tmb_2016}). By combining automatic differentiation with the Laplace approximation, TMB enables scalable maximum likelihood inference in complex SSMs, while recovering unobserved true trajectories. Crucially, our setting differs from standard state-space TMB implementations \citep[e.g.,][]{auger-methe_spatiotemporal_2017} because habitat covariates must be evaluated at unobserved true locations. Specifically, we account for the unobserved covariate values while integrating over the true locations during the optimization process. To our knowledge, this is the first TMB implementation of SSMs handling covariates that depend on latent variables, resolving a fundamental challenge in ecology. The proposed approach is thus fast, intuitive, and powerful, making it well-suited for estimating habitat-selection parameters and UDs from complex tracking data.

We evaluate our method's ability to recover the UD and habitat selection parameters in a simulation study spanning measurement-error and missing-data scenarios, comparing its performance with that of a standard two-step approach. We then illustrate our model with tracking data from narwhal (\textit{Monodon monoceros}), a species vulnerable to the effects of climate change such as the decline in sea ice (\citealp{pizzolato_changing_2014}). Previous studies have shown that bathymetry plays a significant role in narwhal habitat selection (\citealp{laidre2004seasonal,kenyon_baffin_2018, hornby2025behavioural}). In this context, our objective is to assess our method's ability to identify this relationship and to compare it with the results from a two-step approach.

\section{Methods}
\vspace{-0.5cm}
\subsection{Langevin diffusion model for animal movement}
\vspace{-0.3cm}
Movement, represented as the joint velocity and location process, is modeled as a underdamped Langevin diffusion process \citep{michelot_langevin_2019, michelot_multiscale_2024}, which generalizes correlated random walks by allowing movement to be directed towards regions of higher values of a function denoted as $\log\pi$. The stationary distribution $\pi$ represents the (time-invariant) steady-state distribution of the location process. In habitat selection analyses, $\pi$ is typically written as a function of an RSF and corresponds to a UD \citep{michelot_langevin_2019}. Local movement decisions in the `Langevin movement model' reflect habitat selection preferences, with resource availability dynamically constrained by the movement process itself. Since $\pi$ is the limiting distribution of the location process, the long-term pattern of space use stabilizes to $\pi$, creating a direct link between local habitat selection and broad-scale space use.

Let $\boldsymbol{\mu}_t = (\mu_t^x,\mu_t^y)^\top$ and $\boldsymbol{v}_t=(v_t^x,v_t^y)^\top$ denote the location and velocity in the $x$
and $y$ directions at time $t \geq 0$. Given initial conditions $\boldsymbol{\mu}_0, \boldsymbol{v}_0$, their joint dynamics are governed by the following system of stochastic differential equations (SDEs; \citealp{michelot_multiscale_2024}):\\

\begin{equation}
\left\{
\begin{array}{l}
d\boldsymbol{\mu}_t = \boldsymbol{v}_t \, dt, \\
d\boldsymbol{v}_t = -\gamma \boldsymbol{v}_t \, dt + \sigma^2 \nabla \log \pi(\boldsymbol{\mu}_t) \, dt + \sqrt{2 \gamma }\sigma \,d\boldsymbol{B}_t,
\end{array}
\right.
\label{eq:system}
\end{equation}

where $\nabla(\cdot)$ refers to the gradient operator. Within the framework defined by Equation \eqref{eq:system}, the location process $\boldsymbol{\mu}_t$ follows the standard position-velocity relationship, while the equation that drives the velocity process describes more complex patterns that result from three driving forces. First, a friction force $-\gamma \boldsymbol{v}_t$, with $\gamma > 0$, damps the velocity and governs its autocorrelation structure, with large $\gamma$ leading to a rapid decrease of autocorrelation, and low $\gamma$ yielding more persistent movement. Second, a standard Brownian noise $\boldsymbol{B}_t$ captures unexplained variation in the system (e.g., variability in movement and behavior). Third, an environmental drift $\sigma^2 \nabla \log \pi(\boldsymbol{\mu}_t)$ pulls the animal along environmental gradients and towards regions of higher $\pi$, with speed parameter $\sigma$. Since all components of \eqref{eq:system} are isotropic (with scalar $\gamma$ and $\sigma$ and an isotropic gradient), the joint process is isotropic (see \cite{delporte_spatial_2025} for an anisotropic formulation).

Under mild regularity conditions \citep{cheng2018underdamped}, the joint process defined by Equation \eqref{eq:system} has a unique stationary distribution $\pi^{\star}(\boldsymbol{\mu}, \boldsymbol{v}) \propto \pi(\boldsymbol{\mu})e^{-||\boldsymbol{v}||_2^2/2\sigma^2}$ \citep{michelot_multiscale_2024}. Since the location marginal is $\pi$, linking local movement decisions to the long-term UD, while the velocity marginal is a zero-mean Gaussian independent of location and carries no information about habitat selection, we use $\pi$ to refer to the stationary distribution of the location process. Following \cite{michelot_multiscale_2024}, $\pi(\boldsymbol{\mu}) \in \mathbb{R}$ captures how the environment influences animal movement, and is specified as a function of an RSF as follows:
\begin{equation}
\pi(\boldsymbol{\mu}) = \frac{\exp\left( \sum_{k=1}^{K} \beta_k c_k(\boldsymbol{\mu}) \right)}{\int_\Omega \exp\left( \sum_{k=1}^{K} \beta_k c_k(\boldsymbol{\mu^\star}) \right) d\boldsymbol{\mu^\star}},
\label{eq:piCH3}
\end{equation}
where each $c_k$ is a differentiable spatial covariate and $\beta_k$ is its associated selection coefficient. This formulation assumes that, at a broad scale, all locations in the study area are equally accessible, and thus the selection coefficients are interpreted as in a standard RSF \citep{manly2002resource}: positive values of $\beta_k$ indicate preference for higher values of $c_k$, negative values indicate avoidance of higher values of $c_k$ and $\beta_k = 0$ indicates no response. The exponential term in the numerator in Equation \eqref{eq:piCH3} is what is commonly referred to as an RSF \citep{manly2002resource}. The gradient driving the movement process is then $
\nabla \log \pi(\boldsymbol{\mu}) = \sum_{k=1}^{K} \beta_k \nabla c_k(\boldsymbol{\mu}).$ 
The long-term distribution emerging from these local decisions is captured by the stationary distribution $\pi$, which reflects space use at a broad scale. Specifically, $\pi$ corresponds to a UD proportional to an RSF when the availability distribution is uniform across the study area  (Equation \eqref{eq:piCH3}; \citealp{hooten_animal_2017, matthiopoulos2023species}).


We use location data to infer behavior and estimate the dynamics of the unobserved velocity process. Specifically, we are interested in estimating the movement parameters $\sigma$ and $\gamma$, along with the habitat selection parameters $\beta_1,\ldots,\beta_K$. Let $(\boldsymbol{Z}_t)_{t \geq 0} = (\mu_t^x, v_t^x, \mu_t^y, v_t^y)^\top_{t \geq 0}$ denote the joint process of location and velocity. 
Suppose $(\boldsymbol{Z}_t)_{t \geq 0}$ is observed at discrete times $t_0 < t_1 < \cdots < t_n$. We write $\boldsymbol{z}_i = (\mu_i^x, v_i^x, \mu_i^y, v_i^y)^\top$ for the realization of the process at time $t_i$, where $\mu_i^x := \mu_{t_i}^x$ and similarly for the remaining components. The time elapsed between consecutive observations is denoted $\Delta_i = t_{i+1} - t_i$ for $i = 0, \ldots, n-1$. 
 We approximate the joint process of location and velocity by assuming that $\nabla \log{\pi}(\boldsymbol{\mu})$ is constant over the movement path connecting an observation at time $t_i$ with the subsequent observation at $t_{i+1}$.
 With this approximation, $\boldsymbol{Z}_t$ is a Markov process with Gaussian transition densities (\citealp{michelot_multiscale_2024}): 
\begingroup\small
\setlength{\abovedisplayskip}{4pt}
\setlength{\belowdisplayskip}{1pt}
\begin{equation}
\label{discreet}
\boldsymbol{Z}_{t_{i+1}} \mid \{\boldsymbol{Z}_{t_i} = \boldsymbol{z}_i\} \sim \mathcal{N}(\boldsymbol{\eta}_i, \boldsymbol{Q}_i),
\quad 0 \leq i \leq n-1
\end{equation}
\endgroup
with mean vector
\begingroup\small
\setlength{\abovedisplayskip}{3pt}
\setlength{\belowdisplayskip}{3pt}
\begin{equation}
\label{eq:eta_i}
\boldsymbol{\eta}_i =
\begin{pmatrix}
\mu_{i}^x + \dfrac{v_{i}^x(1 - e^{-\gamma \Delta_i})}{\gamma}
+ \dfrac{\sigma^2 \,\partial_x \log \pi(\boldsymbol{\mu}_{i})}{\gamma}
\left( \Delta_i - \dfrac{1 - e^{-\gamma \Delta_i}}{\gamma} \right)
\\[1.2cm]
v_{i}^x e^{-\gamma \Delta_i}
- \dfrac{\sigma^2 \,\partial_x \log \pi(\boldsymbol{\mu}_{i})}{\gamma}
\left(1 - e^{-\gamma \Delta_i}\right)
\\[1.2cm]
\mu_{i}^y + \dfrac{v_{i}^y(1 - e^{-\gamma \Delta_i})}{\gamma}
+ \dfrac{\sigma^2 \,\partial_y \log \pi(\boldsymbol{\mu}_{i})}{\gamma}
\left( \Delta_i - \dfrac{1 - e^{-\gamma \Delta_i}}{\gamma} \right)
\\[1.2cm]
v_{i}^y e^{-\gamma \Delta_i}
- \dfrac{\sigma^2 \,\partial_y \log \pi(\boldsymbol{\mu}_{i})}{\gamma}
\left(1 - e^{-\gamma \Delta_i}\right),
\end{pmatrix}
\end{equation}
\endgroup
 and block diagonal covariance matrix (since $x$ and $y$ evolve independently)
\begingroup\small
\begin{equation}
\label{eq:Q_i}
\boldsymbol{Q}_i =
\begin{pmatrix}
\boldsymbol{Q}_i^{\text{1D}} & \boldsymbol{0} \\
\boldsymbol{0} & \boldsymbol{Q}_i^{\text{1D}}
\end{pmatrix}, \quad
\boldsymbol{Q}_i^{\text{1D}} =
\begin{pmatrix}
\sigma^2\left(\dfrac{2\Delta_i}{\gamma} - \dfrac{e^{-2\gamma\Delta_i}}{\gamma^2} - \dfrac{3}{\gamma^2} + \dfrac{4e^{-\gamma\Delta_i}}{\gamma^2}\right) & \dfrac{\sigma^2}{\gamma}\left(1-2e^{-\gamma\Delta_i}+e^{-2\gamma\Delta_i}\right) \\[0.5cm]
\dfrac{\sigma^2}{\gamma}\left(1-2e^{-\gamma\Delta_i}+e^{-2\gamma\Delta_i}\right) & \sigma^2\left(1-e^{-2\gamma\Delta_i}\right)
\end{pmatrix},
\end{equation}
\endgroup

where $\nabla \log \pi(\boldsymbol{\mu}_i) = (\partial_x \log \pi(\boldsymbol{\mu}_i),\, \partial_y \log \pi(\boldsymbol{\mu}_i))^\top$. This discretization of a continuous-time movement process introduces a bias, which grows as the time interval between observations increases (\citealp{blackwell_joint_2024}). 

\subsection{Measurement model for error-prone observed location data}
\vspace{-0.3cm}
The Langevin movement model traditionally assumes that locations are observed without error. However, most tracking data are recorded with some degree of uncertainty. For example, Argos data are typically accompanied by error estimates derived from least-squares (LS) or Kalman filtering, while Fastloc GPS data are characterized by circular error (\citealp{argos2017, wildlifecomputers_fastloc}). Thus, we couple the Langevin movement model with an observation model that accounts for measurement error, specifically for Kalman-filtered Argos and Fastloc GPS data (\citealp{jonsen_continuous-time_2020}). The core idea is that the observed locations correspond to the true locations $\boldsymbol{\mu}_i$ with additive noise, the magnitude and distribution of which depends on the type of data and the associated uncertainty estimates. Following standard practice in movement ecology, we consider the observed location $\boldsymbol{y}_i$ as a realization of an observation process $\boldsymbol{Y}_i$ with a Gaussian conditional distribution (\citealp{mcclintock_modelling_2015}; \citealp{jonsen_continuous-time_2020}):
\setlength{\abovedisplayskip}{4pt}
\setlength{\belowdisplayskip}{0.1pt}
\begin{equation}
\label{observation_model}
\boldsymbol{Y}_i \mid \boldsymbol{\mu}_i
\sim \mathcal{N}\!\left(\boldsymbol{\mu}_i,\boldsymbol{\Sigma}_i\right).
\end{equation}

For Argos locations, we define $\boldsymbol{\Sigma}_i$ using the error ellipses associated with the location recorded at time $t_i$. Therefore, for a two-dimensional observation process  $\boldsymbol{Y}_{i}$, we use:
\begingroup\small
\begin{center}
\vspace{-1cm}
\begin{equation}
\label{Sigma}
\boldsymbol{\Sigma}_i = \begin{pmatrix}
\sigma_{1,{i}}^2 & \sigma_{12,{i}} \\
\sigma_{21,{i}} & \sigma_{2,{i}}^2
\end{pmatrix}, \qquad \text{with} \qquad
\left\{
\begin{aligned}
\sigma_{1,i}^2 &= \left(\frac{M_i}{\sqrt{2}}\right)^2 \sin^2 r_i + \left(\frac{m_i\psi}{\sqrt{2}}\right)^2 \cos^2 r_i, \\
\sigma_{2,i}^2 &= \left(\frac{M_i}{\sqrt{2}}\right)^2 \cos^2 r_i + \left(\frac{m_i\psi}{\sqrt{2}}\right)^2 \sin^2 r_i, \\
\sigma_{12,i} &= \sigma_{21,i} = \frac{(M_i^2 - m_i^2\psi^2)}{2} \cos(r_i)\sin(r_i),
\end{aligned}
\right.
\end{equation}
\end{center}
\endgroup

where $M_i$ and $m_i$ correspond to the semi-major and semi-minor axis lengths, respectively, $r_i$ corresponds to the orientation of the semi-major axis associated with the observation recorded at time $t_i$ and $\psi$ scales the semi-minor axis of the error ellipse (\citealp{mcclintock_modelling_2015, jonsen_continuous-time_2020}). Thus, Equation \eqref{observation_model} requires that the measurement error associated with each location point (i$.$e$.$, $M_i$ and $m_i$ for all $t_i$) is known and accurate. For Argos data, these are provided with each observation (\citealp{argos2017}). For Fastloc GPS data, we assume circular LS-type measurement error with a $50$m error on each coordinate axis (i.e., $\sigma_{1,i}=\sigma_{2,i}=50$, $\sigma_{12,i} = \sigma_{21,i}=0$), scaled by error-weighting factors $\tau_1$ and $\tau_2$ (see Appendix \ref{Appendix}: Section \ref{app:GPS} for more details). Generally, scaling parameters $\psi, \tau_1$ and $\tau_2$ should be estimated rather than fixed, as this corrects for potential inaccuracies in the reported location uncertainties \citep{jonsen_continuous-time_2020}. Hereafter, we refer to the Langevin movement model with measurement error as the `Langevin SSM'.
\subsection{Inference}
\vspace{-0.3cm}
\label{inferenceCH3}
As both the true positions $\boldsymbol{\mu}_{i}$ and the velocities $\boldsymbol{v}_{i}$ are not observed, they need to be integrated out of the joint likelihood function. Crucially, this means the spatial covariates, which depend on the latent positions, are not observed. We account for this covariate uncertainty by extracting the covariate values $c_k(\boldsymbol{\mu})$ and calculating the gradients $\nabla c_k(\boldsymbol{\mu})$ from raster data during the optimization within TMB (\citealp{kristensen_tmb_2016}), a technique that, to our knowledge, has not been previously used. Because the spatial covariates are a function of the latent locations, they are integrated over all possible true paths via the Laplace approximation.

We conduct inference by maximizing the marginal likelihood function of the observed locations under the Langevin movement model, given by the following expression:
\begin{equation}
    \label{eq:marginal_lik}
    \mathcal{L}(\boldsymbol{\theta}) = \int f_{\boldsymbol{\theta}}(\boldsymbol{y}, \boldsymbol{z}) \; d\boldsymbol{z} = \int g_{\psi,\tau_1,\tau_2}(\boldsymbol{y} \mid \boldsymbol{z}) h_{ \boldsymbol{\beta},\, \sigma,\, \gamma}(\boldsymbol{z}) \; d\boldsymbol{z},
\end{equation}
where $\boldsymbol{\theta} = \left(\psi,\tau_1,\tau_2,\, \boldsymbol{\beta},\, \sigma,\, \gamma\right)$, $\boldsymbol{z} = (\boldsymbol{z}_1, \dots, \boldsymbol{z}_n)$ denotes the stacked latent variables (i.e., true location and velocity process) in the multidimensional case, $\boldsymbol{\beta} = (\beta_1,\ldots, \beta_K)^\top$,
$g_{\psi,\tau_1,\tau_2}(\cdot \mid \boldsymbol{z})$ is the product of Gaussian observation densities defined in Equation~\eqref{observation_model}, 
and $h_{ \boldsymbol{\beta},\, \sigma,\, \gamma}(\cdot)$ is the product of transition densities given in Equation~\eqref{discreet}. Equation~\eqref{eq:marginal_lik} is written for a single track. For $M$ independent tracks sharing model parameter $\boldsymbol{\theta}$, 
the full likelihood is $\prod_{m=1}^{M} \mathcal{L}_m(\boldsymbol{\theta})$, 
where each $\mathcal{L}_m$ takes the same form as Equation \eqref{eq:marginal_lik} 
applied to the $m$-th individual track.

Due to the inherent nonlinearities, the high-dimensional integral is analytically intractable. However, the \textit{Laplace approximation} is a numerically efficient approximate frequentist inference (\citealp{tierney1986accurate}) readily available in TMB (\citealp{kristensen_tmb_2016}; see Appendix \ref{Appendix}: Sections \ref{app:laplace} and \ref{appendixLaplaceImplementation} for more details). The Laplace method approximates the log-integrand by a second-order Taylor expansion (with respect to $\boldsymbol{z}$) centered at its mode, for a given $\boldsymbol{\theta}$. Plugging this approximation back into Equation \eqref{eq:marginal_lik} yields a Gaussian integral with an explicit analytical solution. 
Once the model is fitted, {TMB} provides model parameter estimates and predictions of the random effects (i.e., true locations and velocity). 
The Laplace approximation is exact when the SSM is linear and Gaussian \citep{tierney1986accurate}, a condition violated here by the nonlinearity of $\pi(\cdot)$ in Equation \eqref{eq:eta_i}, making our procedure approximate. Nevertheless, the Laplace approximation often performs well when the integrand $f_{\boldsymbol{\theta}}(\boldsymbol{y}, \boldsymbol{z})$ is dominated by a single mode and the sample size is sufficient to concentrate the distribution (further theoretical details on the Laplace approximation and implementation details are provided in Appendix \ref{Appendix}: Sections \ref{app:laplace} and \ref{appendixLaplaceImplementation}). Our simulation study assesses potential bias introduced by this approximation.

We estimate the UD from the estimated selection coefficients by
$
\hat{\pi}(\boldsymbol{\mu}) \approx 
\frac{\exp\left( \sum_{k=1}^{K} \hat{\beta}_k c_k(\boldsymbol{\mu}) \right)}
{\sum_{j=1}^{M} \exp\left( \sum_{k=1}^{K} \hat{\beta}_k c_k({\boldsymbol{\mu}_j^{\star}}) \right)}, $ where $\{{\boldsymbol{\mu}_1^{\star}}, \ldots, {\boldsymbol{\mu}_M^{\star}}\}$ are locations spanning the study domain $\Omega$, taken as the raster cell centers in practice. Uncertainty in $\hat{\pi}(\boldsymbol{\mu})$ can be quantified from the estimated covariance matrix using the Delta method \citep{oehlert1992note} or Monte Carlo methods \citep{michelot_multiscale_2024}.

\subsection{Simulation}
\vspace{-0.3cm}
We investigated several simulated scenarios to evaluate the performance of our model in estimating both the model parameters and the UD. We also compared its performance with that of a two-step method. Across all scenarios, data are generated from the discretized Langevin 
movement model in Equation~\eqref{discreet} with a constant time interval $\Delta_i = 0.01$, using $5 \times 5000 = 25{,}000$ observations, with a specified proportion of measurement error and missing data. 
Three habitat covariates were simulated with Gaussian random fields and associated selection coefficients $\boldsymbol{\beta} = (-4, 6, 5)$ (see Appendix \ref{Appendix}: Table~\ref{tab:simsettings} for implementation details). Observation noise was added to the simulated locations using Equations ~\eqref{observation_model} and \eqref{Sigma}, 
with semi-major axis $M_i$ and semi-minor axis $m_i$ held constant within each scenario. 
Throughout, $\boldsymbol{\mu}$ denotes the true simulated locations and $\boldsymbol{y}$ the 
noisy observed locations provided as input to each of the compared methods.

We explored different levels of missing data and measurement error (with respect to the speed of the movement). Specifically, we varied the percentage of missing observations (0, 15, 20, 25, 40, 45, 60, 70) under two error conditions: (1) 20\% of error (i.e., $M_i=1$),
and (ii) 40\% of error (i.e., $M_i=2$).
Missing data were assigned completely at random to the simulated track $(\boldsymbol{\mu})$, with the missing locations replaced by \verb|NA| values in the observed track data $(\bm{y})$. When fitting the observed data using TMB, setting missing observations to \verb|NA| (instead of removing entirely) results in the true locations being estimated during periods of missing data. We also explored a range of relative measurement error (1, 10, 20, 25, 30, 40, 50, 60$\%$) without missing data.

To compare the overlap between the estimated UD and the true simulated UD, we use the Bhattacharyya's affinity (BA; \citealp{bhattacharyya1943measure,fieberg_quantifying_2005}):
\setlength{\abovedisplayskip}{4pt}
\setlength{\belowdisplayskip}{4pt}
\begin{equation}
BA = \iint_{-\infty}^{\infty} \sqrt{\vphantom{\widehat{UD}(x,y)}
{UD}_{true}(x,y)}\,\sqrt{\widehat{UD}(x,y)}\,dx\,dy,
\end{equation}
such that the statistic equals $1$ when the two UDs are identical and is $0$ when there is no overlap. We use the BA as a relative measure of similarity between the estimated and true UDs \citep{fieberg_quantifying_2005}: values closer to $1$ indicate better recovery of the true UD, and differences in BA across scenarios reflect differences in estimation accuracy. Intuitively, a higher BA indicates that the two UDs assign similar probabilities of use to the same locations.

The procedure requires initial values for both the model parameters $\boldsymbol{\theta}$ and the latent process $\boldsymbol{z}$ (true locations and velocities). Since complex models can be sensitive to initialization in likelihood-based optimization \citep{auger-methe_state-space_2016}, we adopt a strategy informed by preliminary simulations. Latent locations are initialized at the observed positions, with missing values filled by linear interpolation, and velocities at zero. For the model parameters, we consider two approaches: a neutral initialization, where $\gamma$ and $\sigma$ are estimated from the data and all selection coefficients are set to zero, and an initialization at the true parameter values. To improve numerical stability, an initial optimization over the latent states is performed conditional on fixed model parameters before proceeding to joint optimization over both states and parameters.

For the two-step method, the filtering step fits a correlated random walk with \texttt{aniMotum} as is standard in ecological studies \citep[e.g.,][]{matthews2020killer}. In the second step, we fit the Langevin movement model, treating the predicted locations from the correlated random walk as the ``true'' locations. When data are missing, \texttt{aniMotum} provides predicted locations at both observed and missing time points, effectively imputing the missing data. Thus, the Langevin movement model is fitted on a complete sequence of predicted locations. 
Model fitting is performed using {TMB}. 
\subsection{Narwhal movement data}
\vspace{-0.3cm}
\label{narwhaldatach3}
We demonstrate the practical performance of our method using a narwhal case study from  Qikiqtaaluk, Nunavut, Canada. We focus on estimating habitat selection parameters and the whale's UD. A secondary goal is to enforce spatial constraints and push estimated locations from land into water, rather than discarding or rerouting such points during preprocessing. 

\begin{figure}[H]
    \centering
\caption{Location data of 12 narwhal movement data from the 1st of August to the 2nd of October 2017, colored by individual}    
\includegraphics[width=15cm,height=12cm]{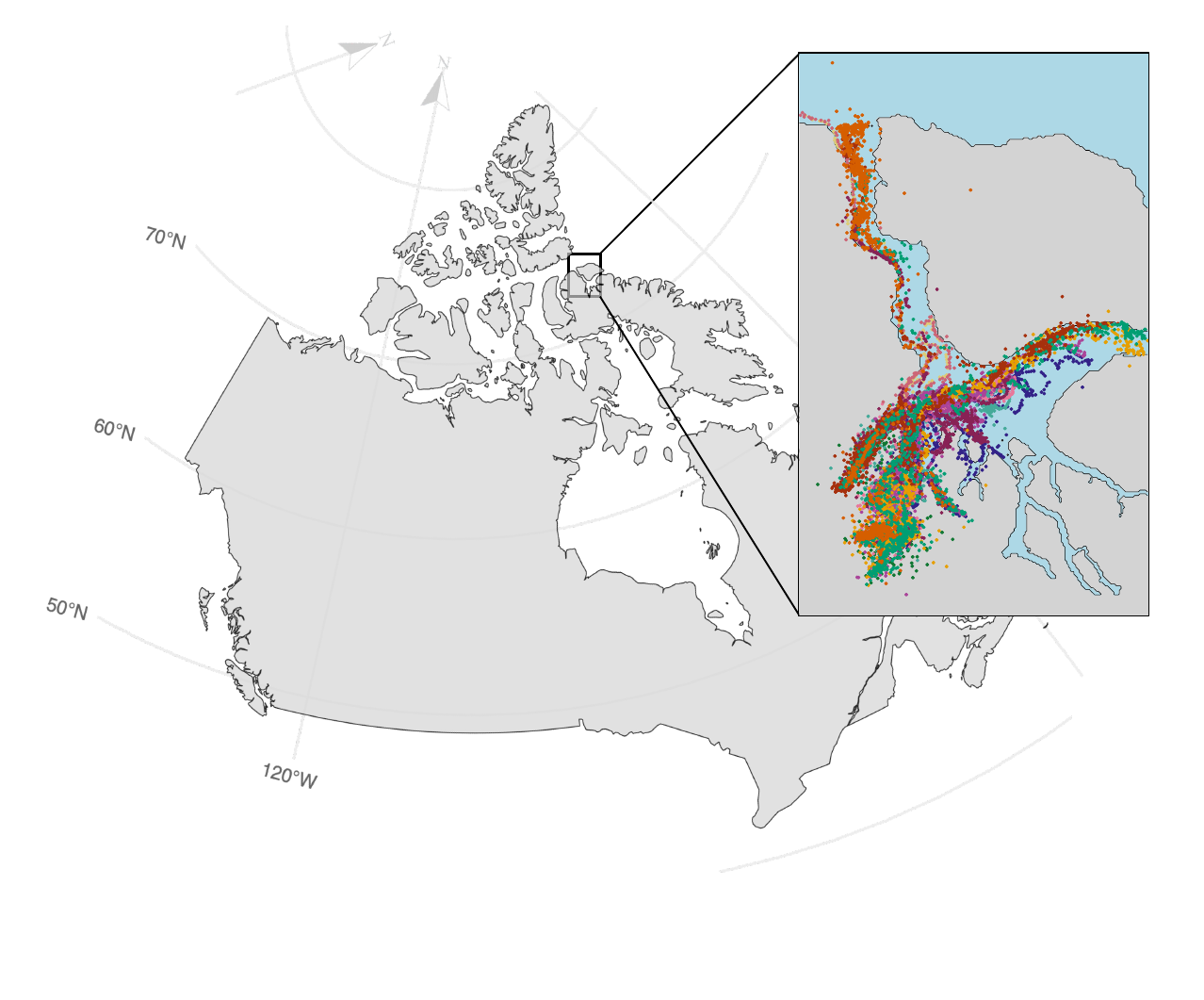}
       \label{fig:narwhaldata}
\end{figure}

We incorporate spatial constraints by including the squared distance to water, $d_{\text{water}}^{2}(\boldsymbol{\mu})$, as an additional covariate into $\pi$, with additional biological knowledge through penalty terms. Using the same $g_{\psi,\tau_1,\tau_2}$ and $h_{\boldsymbol{\beta},\sigma,\gamma}$ 
as in Equation \eqref{eq:marginal_lik}, we get:
\begin{align}
\label{eq:penalized_likelihood}
-\log f_{p,\boldsymbol{\theta}}(\boldsymbol{y}, \boldsymbol{z}) =\ &
-\log g_{\psi,\tau_1,\tau_2}(\boldsymbol{y} \mid \boldsymbol{z}) 
- \log h_{\boldsymbol{\beta},\sigma,\gamma}(\boldsymbol{z})
+ \log p_1(\boldsymbol{z}) \notag \\
& + n\log p_2(\psi) + n\log p_3(\tau_1) + n\log p_4(\tau_2) + n\log p_5(\sigma),
\end{align}
where $p_1$ is a spatial penalty term, $p_2, p_3, p_4$ and $p_5$ are penalty terms on 
the scaling parameters (i.e., $\psi, \tau_1$ and $\tau_2$) and $\sigma$ respectively, and $\boldsymbol{z} = 
(\boldsymbol{z}_1, \dots, \boldsymbol{z}_n)$ are the latent locations 
and velocities treated as random effects. 
We constrained the selection coefficient associated with $d_{\text{water}}^{2}(\boldsymbol{\mu})$ to be negative, encoding the knowledge 
that narwhal avoid land. The spatial penalty term $p_1$ further enforces this 
constraint by penalizing locations increasingly with growing distance from water while assigning no penalty to locations on water:
\vspace{-0.3cm}
\begin{equation}
\log p_1(\boldsymbol{z}) = \lambda\, d_{\text{water}}^{2}(\boldsymbol{\mu}),
\end{equation}
where $\lambda = 10^{5}$. The model parameters $\boldsymbol{\theta}$ are estimated by minimizing 
the Laplace approximation to $-\log\mathcal{L}_p(\boldsymbol{\theta}) = -\log\int f_{p,\boldsymbol{\theta}}(\boldsymbol{y}, \boldsymbol{z})\,d\bm{z}$ (see Appendix \ref{Appendix} Section \ref{app:laplaceNARWHAL} for more details).

The case study and additional simulations (Appendix \ref{Appendix}: Section \ref{app:additional simulations}) showed that including only $p_1$ can cause compensatory parameter inflation such as large $\sigma$ (faster movement, reducing time in penalized habitats) and large $\psi$, $\tau_1$, $\tau_2$ (expanding the ellipse/circle to cover more water). We introduced penalties $p_2$--$p_5$ in Equation \eqref{eq:penalized_likelihood} to prevent such parameter inflation. Penalties on $\psi$, $\tau_1$, and $\tau_2$ were linear:
$
\log p_2(x) = \log p_3(x) = \log p_4(x) = \frac{x}{2},
$
while the penalty on $\log \sigma$ was quadratic:
\begin{equation}
\log p_5(\sigma) = \frac{1}{2}\left(\frac{\log\sigma - \log 4}{0.3}\right)^2.
\end{equation}
An average narwhal speed of $5$ km$\cdot$h$^{-1}$ (\citealp{dietz1995movementsn}) implies $\sigma \approx \sqrt{2/\pi}\times 5 \approx 4$ km$\cdot$h$^{-1}$ under the Langevin model (\citealp{michelot_multiscale_2024}), hence centering on $\log 4$. The scaling of $0.3$ allows deviation from this value while discouraging unrealistic speeds. We do not penalize $\gamma$, as the penalty on $\sigma$ already ensures identifiability (\citealp{michelot_multiscale_2024}). 

In the summer of 2017, 18 narwhal were equipped with tags in Tremblay Sound under approval from the Fisheries and Oceans Canada Animal Care Committee (permit \#AUP 40, S-17/18-1017-NU; \citealp{shuert_decadal_2022}). We evaluate the Langevin SSM using data from the $12$ narwhal with satellite tags recording Fastloc GPS and Argos locations over two months (1st August–2nd October; Figure \ref{fig:narwhaldata}). Tracks with gaps longer than two hours were split into separate tracks (i$.$e$.$, their likelihood contributions are calculated separately), yielding time gaps between consecutive locations ranging from $3.6$ seconds to $2$ hours (average of $15$ minutes). Location uncertainty was handled differently by data type: Kalman-filter-derived error ellipses for Argos data, and a standard $50$ m error radius for Fastloc GPS data. Most Argos observations had large uncertainty (semi-major axis median of $5$ km, third quartile of $15$ km). For the two-step method, \texttt{aniMotum} was used to fit a correlated random walk. For direct comparison, we avoided prefiltering in both methods, using raw locations and disabling \texttt{aniMotum}'s automatic prefiltering.
For both models, we investigated the effect of bathymetry on narwhal habitat selection, as it is an important covariate to explain narwhal movement (\citealp{laidre2004seasonal}; \citealp{kenyon_baffin_2018}; \citealp{hornby2025behavioural}). Bathymetry was obtained from the International Bathymetric Chart of the Arctic Ocean (GEBCO 2024 grid). 

For both methods, we explored $30$ sets of randomly drawn initial values for the habitat selection coefficients to reduce the risk of convergence to a local maximum, sampling $\sigma$, $\gamma$ from $|\mathcal{N}(0,3)|$, while $\psi, \tau_1$ and $\tau_2$ were initialized at one and tracks were initialized at the observed locations. We also considered a neutral initialization, which sets the habitat selection coefficients and velocity to zero and initializes 
$\sigma$ and $\gamma$ from the data. The model yielding the highest likelihood across all $31$ initialization sets ($30$ random and one neutral) was retained.

\section{Results}
\vspace{-0.5cm}
\subsection{Simulation study}
\begin{figure}[H]
\caption{Estimated selection parameters and BA from the Langevin SSM and two-step method. The dashed horizontal line indicates the true parameter value}  
\label{fig:combined}
\vspace{-0.3cm}
    \begin{subfigure}[b]{0.9\textwidth}  
        \hspace{-2cm}
        \includegraphics[width=20.5cm, height=7cm]{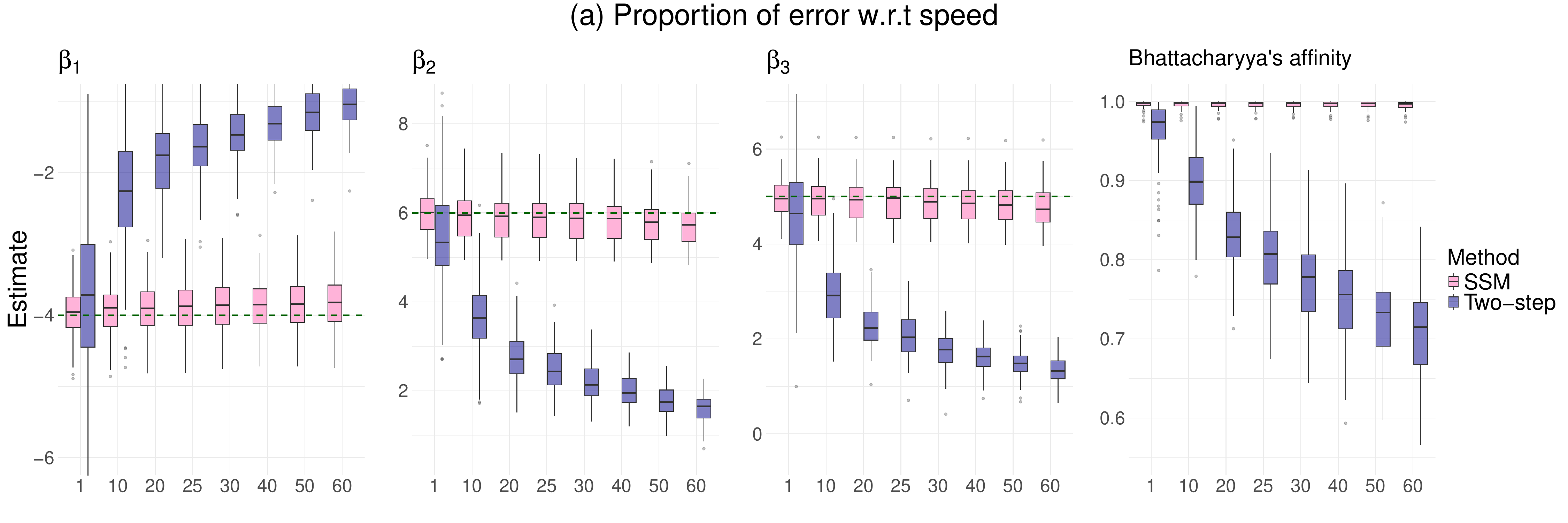}
    \end{subfigure}
     \begin{subfigure}[b]{0.9\textwidth}  
        \hspace{-2cm}
        \includegraphics[width=20.5cm, height=7cm]{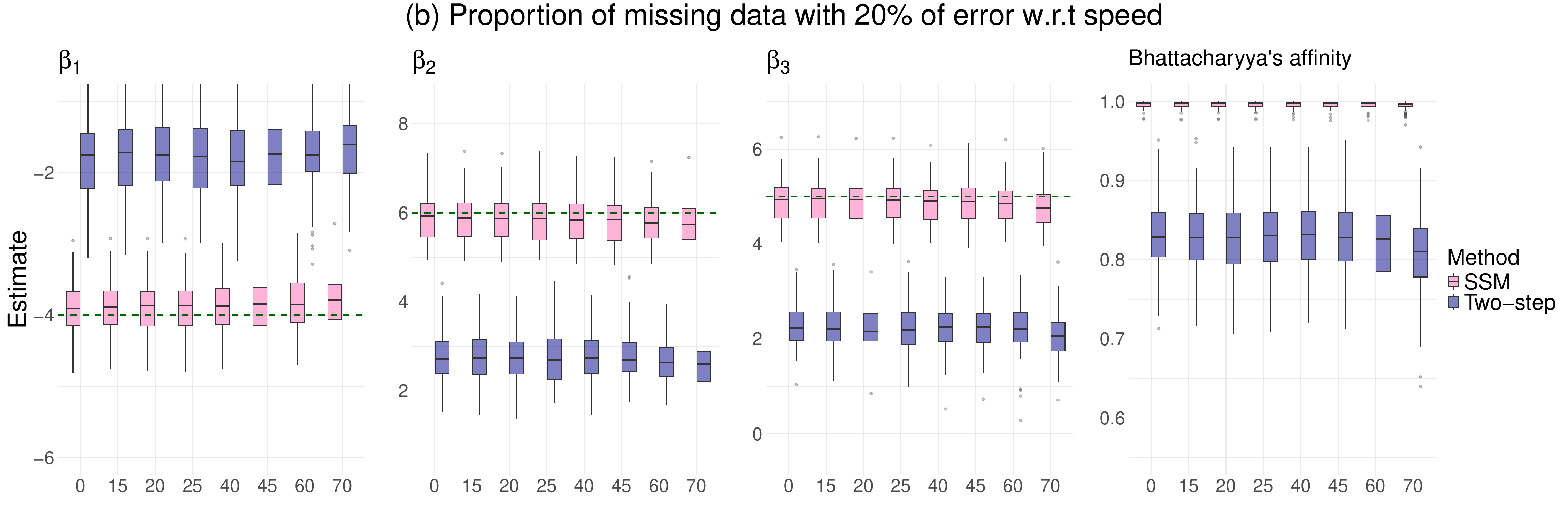}
    \end{subfigure}    
       \begin{subfigure}[b]{0.9\textwidth}  
        \hspace{-2cm}
        \includegraphics[width=20.5cm, height=7cm]{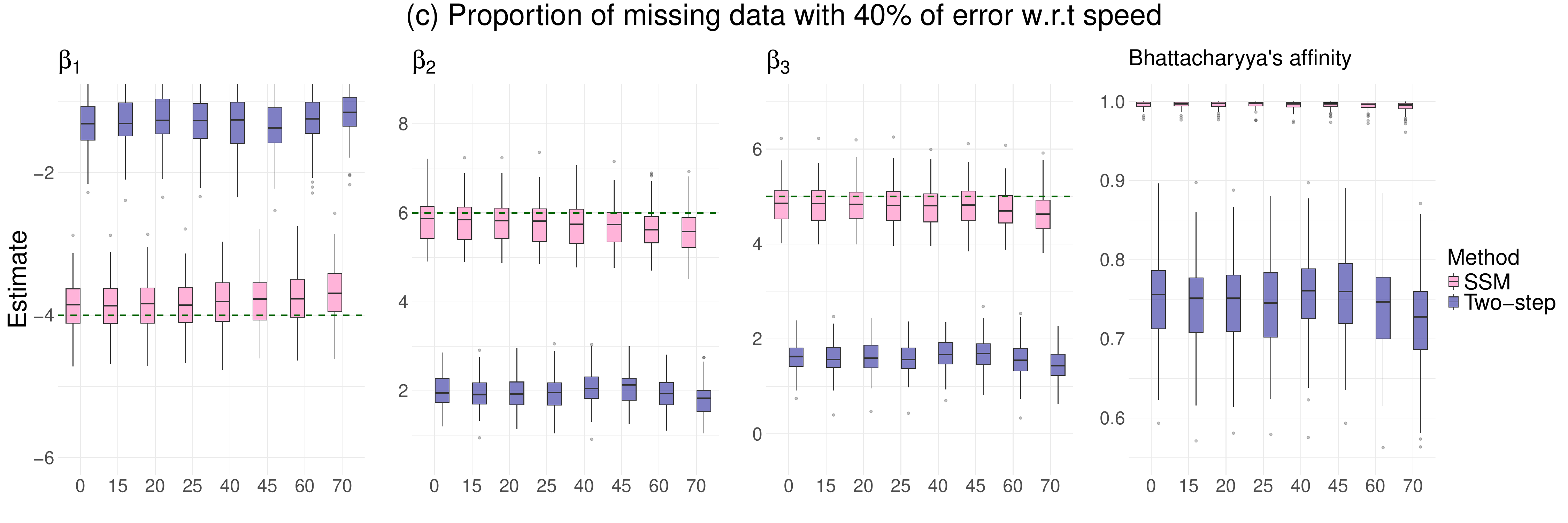}
    \end{subfigure}   
\end{figure}

\vspace{-0.3cm}
\label{SimCH3}
The Langevin SSM outperforms the two-step method in estimating parameters for both movement and habitat selection (Figure \ref{fig:combined}). The Langevin SSM consistently maintains a high Bhattacharyya’s affinity (BA $\approx$ 1), demonstrating its strong performance in preserving space-use information, even with large measurement error in the location data. In all scenarios, the Langevin SSM yields substantially less biased estimates of habitat selection parameters, whereas the two-step approach systematically underestimates the strength of habitat selection. 
The Langevin SSM recovers trajectories nearly identical to the true (simulated) ones, while the two-step method still carries much of the added noise (Appendix \ref{Appendix}: Figure \ref{fig:tracks_both2}). This residual noise also impacts the underlying movement-habitat relationship, leading the two-step approach to inflate the movement parameter estimates (Appendix \ref{Appendix}: Figure \ref{fig:gammasigma}). 

Measurement error is the main source of bias for both methods (Figures \ref{fig:combined}a-c). As measurement error increases, estimated selection parameters in both methods are increasingly biased towards zero. This bias is severe for the two-step method even at low error levels, while it grows more gradually for the Langevin SSM. This bias likely occurs because larger measurement errors degrade the signal in the data, making it harder to capture true habitat selection patterns. 
However, our method exhibits substantially less bias than the two-step approach. With 50\% measurement error, the two-step bias reaches $-3.53$ for $\beta_3$ (i$.$e$.$, $60\%$ of the true value) compared to $-0.15$ ($2.5$\%) for the Langevin SSM (Appendix \ref{Appendix}: Tables \ref{tab:simulation_results2}-\ref{tab:simulation_results4}). Moreover, beyond 20\% measurement error, the two-step method achieves zero coverage (95\% CIs) for all selection parameters, while the Langevin SSM maintains coverage between $0.85$ and $0.97$. In scenarios with 20\% and 40\% measurement error and increasing missing data, the Langevin SSM outperforms the two-step method in terms of bias and BA (Figure \ref{fig:combined}b-c). The two-step method consistently yields zero coverage, while the Langevin SSM maintains more satisfying coverage (81\%-97\% Appendix \ref{Appendix}: Tables \ref{tab:simulation_results2}-\ref{tab:simulation_results4}). However, both methods remain robust to missingness, with bias staying stable.  

\subsection{Case study}
\vspace{-0.3cm}
\label{NarwhalCH3}

\begin{figure}[H]
\caption{Left: bathymetry raster from the International Bathymetric Chart of the Arctic Ocean \citep{gebco2024}. Middle: logarithm of the estimated utilization distribution from the two-step model. Right: logarithm of the estimated utilization distribution based on the Langevin SSM.
}  
\hspace{-2.5cm}
\includegraphics[width=1.3\textwidth]{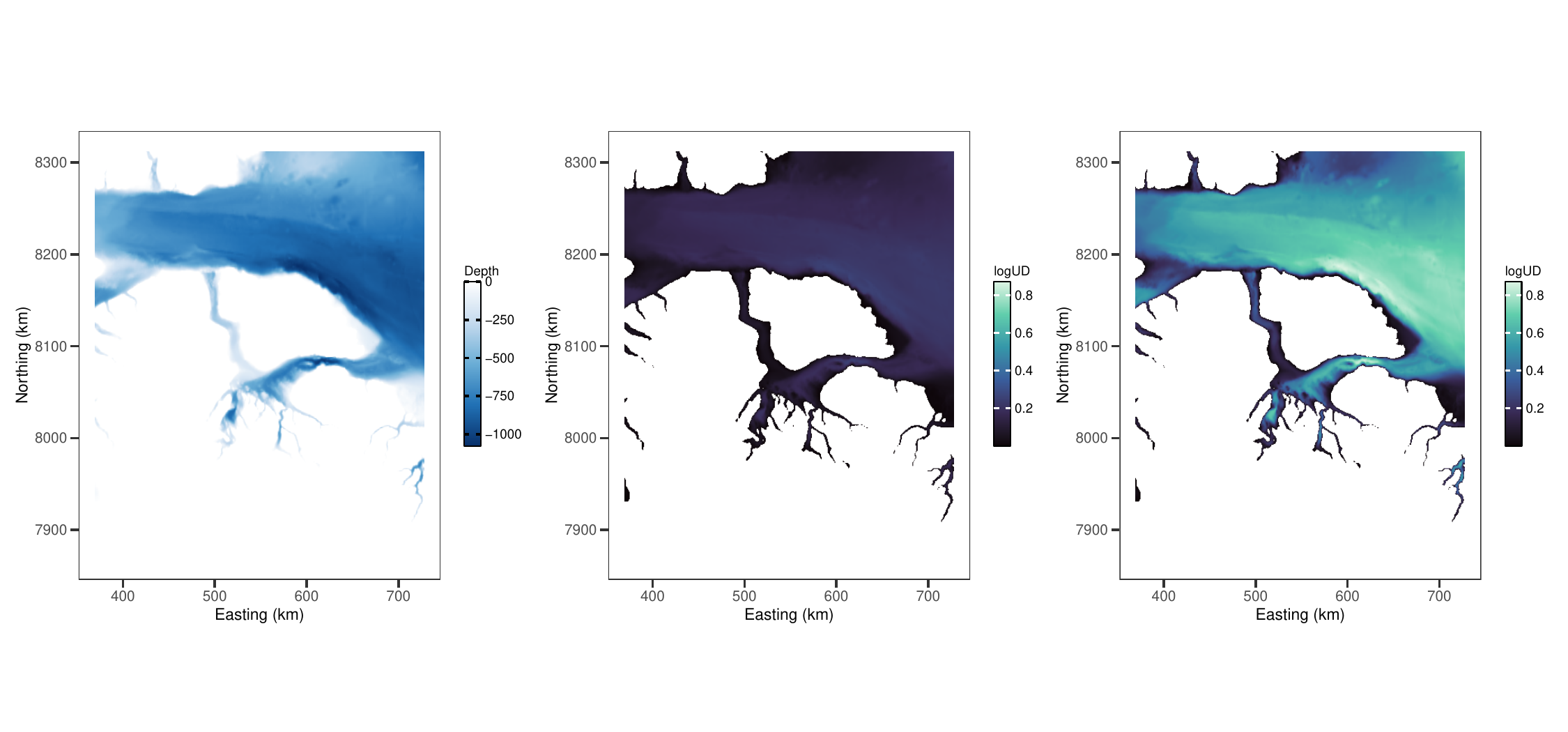}
\label{fig:langevin_ud}
\end{figure}

The Langevin SSM identifies bathymetry as an important driver of narwhal habitat selection, with an estimated coefficient $\hat{\beta}_1= -4.3$ (95\% CI $[-5.5, -3]$), suggesting a preference for deeper waters. The two-step approach yields a weaker estimate ($\hat{\beta}_{1,\text{two-step}} = -1.3$, 95\% CI $[-1.8, -0.7]$). The habitat selection estimates are consistent with the simulation results, which show that the two-step method tends to underestimate covariate effects on habitat selection in the presence of measurement error, while the Langevin SSM is more reliable. The wider confidence intervals of the Langevin SSM reflect the propagation of location uncertainty into parameter estimation, while the two-step approach ignores this uncertainty, leading to narrower intervals. In Qikiqtaaluk, the areas with the 10\% highest estimated UD from the Langevin SSM on the log scale (i.e., the top 10\% of log-UD values), correspond to depths lower than $900$ meters (Figure \ref{fig:langevin_ud}). 

The estimated speed parameter from the Langevin SSM is $5.2$, which corresponds to an average movement speed of $6.5$km$.$h$^{-1}$ (\citealp{michelot_multiscale_2024}). In contrast, the two-step approach yields an implausibly large value of $\hat{\sigma}_{\text{ two-step}} = 384{,}406$ (i.e., $481{,}781$ km$.$h$^{-1}$). A similar discrepancy is observed for the persistence parameter. 
The Langevin SSM model estimates $\hat{\gamma} = 11$ (and $\log(\hat{\gamma}_{\text{ two-step}}) \approx 23$), leading to a persistence time scale of approximately 16 minutes (calculated as $3/\hat{\gamma}$), which reflects the duration over which the velocity autocorrelation decreases by about 95\% (\citealp{michelot_multiscale_2024}). This relatively low persistence velocity could be explained by the intricate fjord system, which prevents narwhal from maintaining straight trajectories for prolonged periods. The extreme discrepancy between the two methods' estimates mirrors the patterns observed in the simulation study, where the two-step method yields exploding estimates for both the speed and persistence parameters, making the results unreliable. 

The raw location data contain $3{,}722$ unrealistic locations on land. The two-step did not lead to any substantial improvement, with $3{,}721$ unrealistic locations on land. In contrast, the Langevin SSM successfully pushed a substantial number of these points into water (Appendix \ref{Appendix}: Figure \ref{fig:langevin_locations_panels}c). Specifically, $2{,}825$ locations were pushed to water and $191$ from water to land, resulting in a total of $1{,}088$ points left on land. 

\section{Discussion}
\vspace{-0.3cm}
We proposed a Langevin state-space model to infer animal habitat selection and utilization distributions from irregular, error-prone, and complex movement data. Maximum likelihood inference is carried out efficiently using the Laplace approximation and automatic differentiation as implemented in TMB, with an R package for simulating and fitting the model available on GitHub (\url{github.com/bmcclintock/langevinSSM}).
The proposed method greatly improves upon the traditional two-step approach, which consists of first fitting a simpler model to account for measurement error and/or temporal irregularity and then fitting a more complex model to the predicted locations from the first stage. Our simulations showed that the Langevin SSM better preserves the UD, minimizes bias in habitat selection parameters, and improves the recovery of individual trajectories. When applied to narwhal data, our method preserves more of the information contained in the raw observations than the two-step approach, enabling the recovery of a stronger effect of bathymetry. Narwhal tracking data are highly irregular, combine two data sources (Fastloc GPS and Argos), and are collected in complex landscapes such as fjord systems. The strong performance of our method reflects its suitability for accommodating these data complexities.

As measurement error increases, both methods tend to underestimate the strength of habitat selection, with parameter estimates increasingly biased towards zero selection. This bias reflects a loss of information from greater uncertainty, weakening the habitat-selection signal. However, this bias is consistently larger and more substantial for the two-step method, even with low measurement error, while it grows only gradually with the Langevin SSM. This discrepancy in bias suggests that two-step approaches are not well suited to accommodate substantial measurement error. By remaining robust to such uncertainty, our method provides a more reliable framework for estimating UDs and habitat-selection parameters. Both methods appeared robust to missing data, likely because they are formulated in continuous time and evaluated on a fine temporal grid, such that sporadic removals of observations have only a negligible impact on inference. While we examined a broad range of scenarios, future research should investigate the effects of sampling rate and covariate correlation on the performance of our method. Coarser sampling intervals may introduce bias (e.g., \citealt{mcclintock_multistate_2024,michelot_multiscale_2024}, but see \citealt{blackwell_joint_2024} for an approach to mitigating such attenuation bias), while correlated covariates can reduce inferential power. Our available code and the speed with which our model can be fitted facilitate further simulation studies. For example, simulations with $n=25{,}000$ observations on average required 14 sec per model fit on an Apple M2 Max with 64GB RAM.

The relatively poor performance of the two-step method was expected, since the true data-generating model was a Langevin diffusion, while the two-step method's first stage fits a correlated random walk. However, it is common practice in movement ecology to filter with a simpler model before fitting a more complex model to the processed tracks for habitat selection analysis. In principle, the two-step approach could be improved with a more flexible first-stage model, such as a move persistence model (\citealp{auger-methe_spatiotemporal_2017}), which extends the correlated random walk by allowing behavioral persistence to vary over time. However, upgrading from a simple to a correlated random walk did not greatly improve performance, with comparable bias in the selection coefficients and inflation in $\gamma$ and $\sigma$ (Appendix \ref{Appendix}: Section \ref{app:additional results}). Together, these results suggest that the core limitation of the two-step approach lies not in the choice of the movement model but in the decoupling of movement from habitat selection. Since habitat selection inherently influences movement, estimating movement independently in a first step necessarily introduces bias. Jointly estimating both, as in the Langevin SSM, better accounts for measurement error and temporal irregularity in the data.

Our results suggested an increased use of deep-water habitats, consistent with previous studies (\citealp{richard1994distribution}; \citealp{npc2000}), and aligning with \cite{watt_spatial_2017} who estimated a summer kernel utilization distribution focused along the deep central section of Eclipse Sound. As deep-diving marine mammals, narwhal likely use these areas primarily for foraging (\citealp{watt_spatial_2017}). Deep diving could also reflect responses to disturbance (\citealp{dupont2026estimating}), or socializing at the surface over deep waters (\citealp{watt_spatial_2017}; \citealp{hornby2025behavioural}). While our model assumes a linear relationship with bathymetry, the true relationship could be non-linear, with narwhal selecting a depth range rather than responding uniformly across the bathymetric gradient. Additional covariates, such as distance to shore and seafloor slope, are also expected to influence habitat selection (\citealp{hornby2025behavioural}).

After developing our approach, we became aware of related work by \cite{delporte_spatial_2025}, who independently extend the Langevin model to handle measurement error via an SSM and enforce spatial constraints within the SDE, though their focus is on track reconstruction and their fitting procedure differs considerably from ours. 
While their method effectively reconstructs true trajectories, it uses particle filtering with known model parameters and does not explore habitat selection inference. 
Building on the same modeling framework, we express the stationary distribution as an RSF and jointly estimate all model parameters and latent states using TMB for faster inference. Our penalty term on the negative log-likelihood and the use of $d_{\text{water}}^{2}$ as a covariate in $\pi(\boldsymbol{\mu})$ jointly discourage locations from being estimated on land. The latter is closely related to the spatial-constraint approach of \cite{delporte_spatial_2025}, and differences in performance are likely attributable to different modeling choices (see Appendix \ref{Appendix} Section \ref{app:additional results} for details).


We did not explicitly address the bias that can be introduced by the discretization in equation \eqref{discreet}, which assumes that $\nabla \log\pi(\boldsymbol{\mu})$ remains constant along the movement path between consecutive observations. 
This bias should not affect our simulation study, since we use a fine temporal grid, but it may be problematic in applications to real data with irregular and sometimes large observation gaps. 
Practitioners working with coarser temporal resolutions (e.g., a couple of observations a day) should consider the method proposed by \cite{blackwell_joint_2024} to reduce this bias. Our framework can accommodate their method, but at an increased computational cost. 
\cite{delporte_spatial_2025} tackle this discretization bias 
using splitting schemes. 
However, their method would also likely increase computational cost. A further limitation is the Gaussian observation model for Argos data, which may not capture heavy-tailed error distributions \citep{delporte_spatial_2025} or the temporal dependence induced by Argos's error-correction processing. Future work could therefore explore more flexible alternatives, such as a Student's t-distribution or an autoregressive error structure.

The Langevin movement model provides a natural and powerful framework for quantifying habitat selection and deriving utilization distributions from spatial covariates using locations collected at irregular time intervals. By extending the Langevin movement model to account for measurement error, we provide a tool that can be used to tackle the main challenges left in applying habitat selection models to error-prone data.

\section*{Acknowledgments}
\vspace{-0.3cm}
We thank the Natural Sciences and Engineering Research Council of Canada (NSERC), Canada Research Chairs program, BC Knowledge Development Fund and Canada Foundation for Innovation's John R. Evans Leaders Fund, Canadian Statistical Sciences Institute (CANSSI), Fisheries and Oceans Canada (DFO), Arctic Section of the Society of Naval Architects and Marine Engineers, Polar Continental Shelf Program, Nunavut Wildlife Management Board, Nunavut Implementation Fund, and World Wildlife Fund Canada for their support. We thank the community of Mittimatalik (Pond Inlet) for its support in tagging operations and the devoted people who led operations in the field. We are grateful to Dr$.$ Théo Michelot for the constructive discussions. The findings and conclusions in the manuscript are those of the authors and do not necessarily represent the views of the National Marine Fisheries Service, NOAA. Any use of trade, product, or firm names does not imply an endorsement by the US Government.
\section*{Author Contributions}
\vspace{-0.3cm}
BM, MAM, and FD conceived the study; BM developed the code, simulation framework, and R package; NH and MM conducted fieldwork; FD prepared the data, contributed to model-fitting and computational implementation, performed the analyses, and led the writing; MAM and MM supervised FD; JOF assisted with computational implementation and writing. All authors provided feedback on drafts and approved the final version.

\section*{Conflict of Interest Statement}
\vspace{-0.3cm}
The authors declare no conflict of interest.

\putbib
\newpage

\end{bibunit}

\newpage
\appendix
\begin{bibunit}
\setcounter{section}{0}
\renewcommand{\thesection}{S\arabic{section}}
\renewcommand{\thesubsection}{S\arabic{subsection}}
\renewcommand{\thetable}{S\arabic{table}}
\renewcommand{\thefigure}{S\arabic{figure}}
\renewcommand{\theequation}{S\arabic{equation}}
\setcounter{table}{0}
\setcounter{figure}{0}
\setcounter{equation}{0}

\refstepcounter{section}
\begin{center}
{\Large\bfseries Appendix S1}\\[1em]
{\Large Inferring resource selection and utilization distributions from irregular and error-prone animal tracking data
}\\[1.5em]
{\normalsize
Fanny Dupont,
Brett T.\ McClintock,
Jan-Ole Fischer,
Marianne Marcoux,\\
Nigel E.\ Hussey,
Marie Auger-Méthé
}
\end{center}
\vspace{2em}
\label{Appendix}
\subsection{Observation model for Fastloc GPS data}
\label{app:GPS}
Recall that the observed location $\boldsymbol{y}_i$ is a realization of an observation process $\boldsymbol{Y}_i$ centered on the true location $\boldsymbol{\mu}_i$, with a Gaussian conditional distribution:
\setlength{\abovedisplayskip}{4pt}
\setlength{\belowdisplayskip}{0.1pt}
\begin{equation}
\label{observation_model:GPS}
\boldsymbol{Y}_i \mid \boldsymbol{\mu}_i
\sim \mathcal{N}\!\left(\boldsymbol{\mu}_i,\boldsymbol{\Sigma}_i\right).
\end{equation}
For Fastloc GPS data, we assume circular LS-type measurement error with a baseline error of $50$m on each coordinate axis (i.e., $\sigma_{1,i}=\sigma_{2,i}=50$ and $\sigma_{12,i} = \sigma_{21,i}=0$), and introduce error-weighting factors $\tau_1$ and $\tau_2$ that scale this baseline error separately for each direction (\citealp{jonsen_continuous-time_2020}; \citealp{wildlifecomputers_fastloc}), giving the following diagonal covariance matrix:

\begin{equation}
\label{Sigma_fastloc}
\boldsymbol{\Sigma}_i = \begin{pmatrix}
(\sigma_{1,i}\tau_1)^2 & 0 \\
0 & (\sigma_{2,i}\tau_2)^2
\end{pmatrix}.
\end{equation}
\subsection{Mathematical details of the joint negative log-likelihood}
\label{mathdetCH3}
In this section, we detail the joint negative log-likelihood (i$.$e$.$, the negative log-integrand in Equation \eqref{eq:marginal_lik}  to provide further insight into the structure of the model. Let $\boldsymbol{y} = (\boldsymbol{y}_0, \ldots, \boldsymbol{y}_n)$ denote the vector of observations at times $t_0, \ldots, t_n$, and let
$\boldsymbol{z} = (\boldsymbol{z}_0, \ldots, \boldsymbol{z}_n)$ denote the associated vector of latent states,
where $\boldsymbol{z}_i$ comprises location and velocity for $0 \leq i \leq n$. Then we have:

\begin{center}
\begin{equation}
    \begin{aligned}
\log (g_{\psi,\tau_1,\tau_2}(\boldsymbol{y} \mid \boldsymbol{z}) h_{ \boldsymbol{\beta},\, \sigma,\, \gamma}(\boldsymbol{z}))
&= \sum_{i=0}^{n} \log g_{\psi,\tau_1,\tau_2}(\boldsymbol{y}_i \mid \boldsymbol{z}_i)
   + \sum_{i=0}^{n-1}\log  q(\boldsymbol{z}_{i+1};\boldsymbol{\eta}_{i}, \boldsymbol{Q}_i), \\
&= \sum_{i=0}^{n} \log q(\boldsymbol{y}_i; \boldsymbol{\mu}_i, \boldsymbol{\Sigma}_i)
   +\sum_{i=0}^{n-1} \log q(\boldsymbol{z}_{i+1};\boldsymbol{\eta}_{i}, \boldsymbol{Q}_i),
\end{aligned}
\end{equation}
\end{center}
where $q(\cdot; \boldsymbol{\mu}, \boldsymbol{\Sigma})$ denotes the probability density function of a multivariate Gaussian distribution with mean vector $\boldsymbol{\mu}$ and covariance matrix $\boldsymbol{\Sigma}$. The observation model covariance matrix $\boldsymbol{\Sigma}_i$ is defined in Equation \eqref{Sigma} for Argos location data and Equation \eqref{Sigma_fastloc} for Fastloc GPS data, and
$\boldsymbol{\eta}_i$ and $\boldsymbol{Q}_i$ are defined in
Equations \eqref{eq:eta_i} and \eqref{eq:Q_i}, respectively.

\subsection{Theoretical background on the Laplace approximation}
\label{app:laplace}
In this section, we provide further details on the theoretical foundations of the Laplace approximation (\citealp{tierney1986accurate, erkanli1994laplace,vandervaart}) and discuss potential sources of bias.

The Laplace approximation assumes that the integrand 
$f_{\boldsymbol{\theta}}(\boldsymbol{y}, \boldsymbol{z})$, viewed as a 
function of $\boldsymbol{z}$, is well approximated by a Gaussian centred 
at its mode \citep{tierney1986accurate}. Since \linebreak $g_{\psi,\tau_1,\tau_2}(\boldsymbol{y} 
\mid \boldsymbol{z})$ is Gaussian in $\boldsymbol{z}$ by 
Equation~\eqref{observation_model}, the only source of non-Gaussianity 
in the integrand is the density 
$h_{\boldsymbol{\beta},\sigma,\gamma}(\boldsymbol{z})$.  When the 
transition function between consecutive latent variables is Gaussian and linear in $\boldsymbol{z}$, the integrand is exactly Gaussian and the Laplace approximation is exact 
\citep{tierney1986accurate}. However, the nonlinearity introduced by 
$\pi(\cdot)$ in Equation~\eqref{discreet} renders the 
integrand non-Gaussian in $\boldsymbol{z}$, which can introduce bias in 
the approximated marginal likelihood. Nevertheless, the Laplace approximation often performs remarkably well in practice, particularly when the integrand $f_{\boldsymbol{\theta}}(\bm{y},\bm{z})$ is dominated by a single mode and the sample size provides sufficient information to concentrate the distribution (\citealp{tierney1986accurate}). While the approximation generally 
fails when the integrand has multiple modes, a Gaussian approximation 
can remain appropriate even when the integrand is non-Gaussian, provided 
it is unimodal and well-concentrated around its mode 
\citep{bruijn1961asymptotic, tierney1986accurate}. 

An additional theoretical consideration is that standard asymptotic justifications for the Laplace approximation do not apply when the number of random effects grows with the sample size \citep{rue2009approximate}, as is the case here since both locations and velocities are treated as latent variables. However, the Markov structure implies that each latent state $\boldsymbol{z}_i$ depends only on its neighbors and its associated observation $\boldsymbol{y}_i$, such that the approximation relies on the local rather than global behavior of the integrand. Thus the Laplace approximation is expected to perform well when time intervals $\Delta_i$ are short, such that $\nabla\log\pi$ varies little between consecutive observations and the transition density between consecutive latent states is close to Gaussian. We therefore recommend limiting large time gaps in the data and performing track segmentation when necessary (e.g., as described in 
Section~\ref{narwhaldatach3}).

\subsection{Additional implementation details on the Laplace approximation}
\label{appendixLaplaceImplementation}
Conveniently, the Laplace approximation is fully automated in the {TMB} \texttt{R} package (\citealp{kristensen_tmb_2016}), requiring only that the user provides the negative log-integrand from Equation~\eqref{eq:marginal_lik} as a \texttt{C++} script. The package then returns the (negative) logarithm of the Laplace approximation to Equation~\eqref{eq:marginal_lik} along with its gradient as a standard \texttt{R} function, which can be directly used for numerical optimization with quasi-Newton methods. During each optimization step—i.e., each call to the function or its gradient—{TMB} automatically performs an \textit{inner} optimization over $\boldsymbol{z}$, required for the Taylor expansion in the Laplace approximation.
For users familiar with \texttt{C++}, {TMB} provides a flexible framework to implement fast, automatic Laplace approximations for a wide range of models, requiring only the specification of the negative log-likelihood (\citealp{kristensen_tmb_2016}; \citealp{auger-methe_spatiotemporal_2017}; \citealp{augermethe_guide_2021}; \citealp{mcclintock_worth_2021}; \citealp{breivik2025spatial}).
Once the model is fitted, TMB provides model parameter estimates and, conveniently, predictions of the random effects (true locations and velocity in our case) as well as uncertainties for both, respectively. The package is numerically highly efficient by using automatic differentiation to compute first and second derivatives required for the Laplace approximation, as well as by automatically detecting sparsity in the second derivative function.

\subsection{Additional details on the Laplace approximation in the narwhal case study}
\label{app:laplaceNARWHAL}
In this section, we provide details on the Laplace approximation used 
to evaluate the penalized marginal likelihood 
$\mathcal{L}_p(\boldsymbol{\theta})$ in the narwhal case study.

The model parameters 
$\boldsymbol{\theta}$ are estimated by minimizing the Laplace 
approximation to $-\log\mathcal{L}_p(\boldsymbol{\theta}) = 
-\log\int f_{p,\boldsymbol{\theta}}(\boldsymbol{y}, \boldsymbol{z})\, 
d\boldsymbol{z}$, given by: 
\begin{equation}
-\log \mathcal{L}_p(\boldsymbol{\theta}) \approx 
-n\log\sqrt{2\pi} 
+ \frac{1}{2}\log\det\mathbf{H}(\boldsymbol{\theta}) 
- \log f_{p,\boldsymbol{\theta}}(\boldsymbol{y}, \hat{\boldsymbol{z}}(\boldsymbol{\theta})),
\end{equation}
where $\hat{\boldsymbol{z}}(\boldsymbol{\theta})$ is the mode of 
$-\log f_{p,\boldsymbol{\theta}}(\boldsymbol{y}, \boldsymbol{z})$ with 
respect to $\boldsymbol{z}$ for fixed $\boldsymbol{\theta}$, and 
$\mathbf{H}(\boldsymbol{\theta})$ is the Hessian of $-\log 
f_{p,\boldsymbol{\theta}}(\boldsymbol{y}, \boldsymbol{z})$ with respect 
to $\boldsymbol{z}$, evaluated at $\hat{\boldsymbol{z}}(\boldsymbol{\theta})$.

In practice, this approximation is fully automated by {TMB} 
\citep{kristensen_tmb_2016} and the user supplies the negative 
log-integrand $-\log f_{p,\boldsymbol{\theta}}(\boldsymbol{y}, 
\boldsymbol{z})$ as a \texttt{C++} function, and {TMB} 
internally computes the mode $\hat{\boldsymbol{z}}(\boldsymbol{\theta})$, 
the Hessian $\mathbf{H}(\boldsymbol{\theta})$, and returns 
$-\log\mathcal{L}_p(\boldsymbol{\theta})$ along with its gradient 
with respect to $\boldsymbol{\theta}$.

\subsection{Details on the simulation setting}
\begin{table}[H]
\centering
\caption{Simulation setting across all scenarios}
\label{tab:simsettings}
{\fontsize{10}{12}\selectfont
\begin{tabular}{lll}
\hline
\textbf{Parameter} & \textbf{Value} & \textbf{Description} \\
\hline
\texttt{nsims} & 100 & Number of simulation runs \\
\texttt{nbAnimals} & 5 & Number of tracks per simulation \\
\texttt{obsPerAnimal} & 5000 & Observations per track \\
{$\Delta_i$} & 0.01 & Simulation time step \\
{$\boldsymbol{\beta}$} & c(-4, 6, 5) & Selection coefficients for cov$_1$, cov$_2$, cov$_3$ \\
{$\sigma$} & 5 & Speed parameter of the Langevin movement model \\
{$\gamma$} & 0.5 & Friction parameter \\
{$\psi$} & 1 & Measurement-error scaling \\
\texttt{propMissing} & varies & Proportion of missing observations \\
{$M_i$} & varies (constant within each scenario) & SD of major axis of error ellipse \\
{$m_i$} & $M_i/2$ & SD of minor axis \\
{$r_i$} & c(0, 180) & Error-ellipse orientation (degrees) \\
\texttt{covRange} & c(0.1, 0.5) & Covariate spatial range \\
\hline
\end{tabular}
}
\end{table}

\subsection{Additional results}
\label{app:additional results}
\subsection*{Simulation study}
This section provides additional figures and tables from the simulation study.
\newpage
\begin{figure}[H]
    \caption{Histogram of Langevin SSM estimates for $\beta_1$, $\beta_2$, and $\beta_3$ across varying percentage of error with respect to movement speed, over $100$ simulated datasets. The percentages are given at the top. The red lines indicate the true parameter values.}
    \hspace{-1cm}
    \label{fig:tracks_both1}
\includegraphics[width=18cm,height=6.5cm]{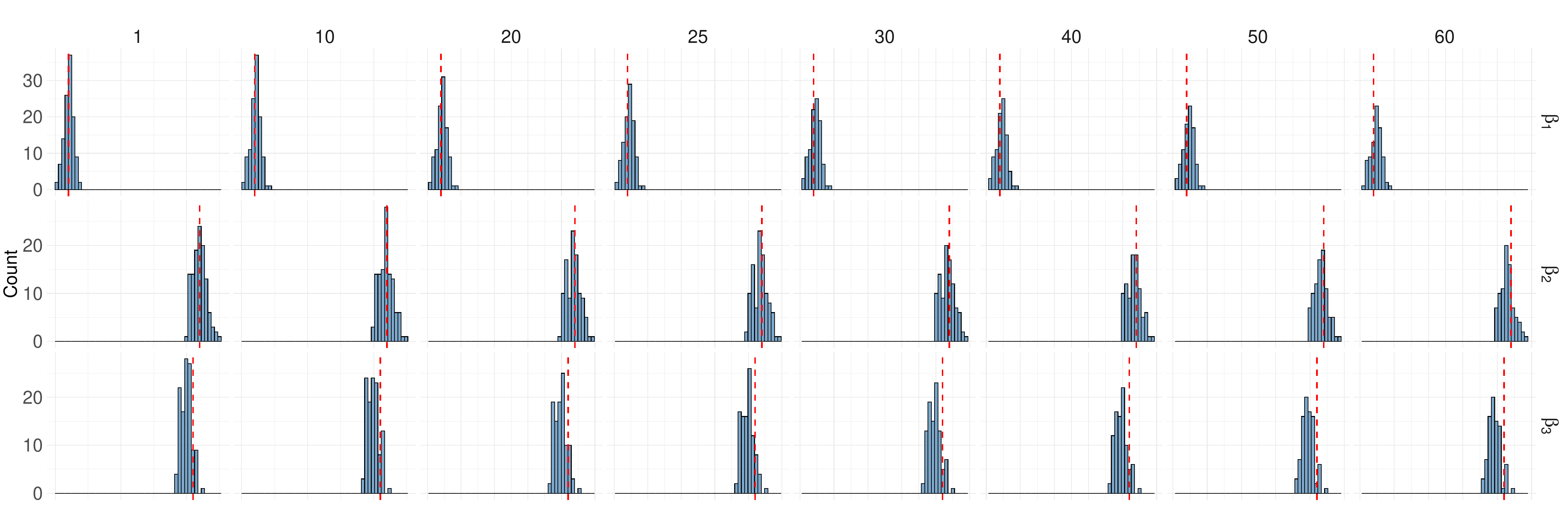} 
\end{figure}
\begin{figure}[H]
    \caption{Example of tracks estimated using the Langevin SSM and the two-step method. The simulated true tracks (green) are shown without error, while 50\% location error was added to the simulated tracks before model fitting.}
    \label{fig:tracks_both2}
\vspace{3cm}
    \hspace{-1cm}
\includegraphics[width=18cm,height=6cm]{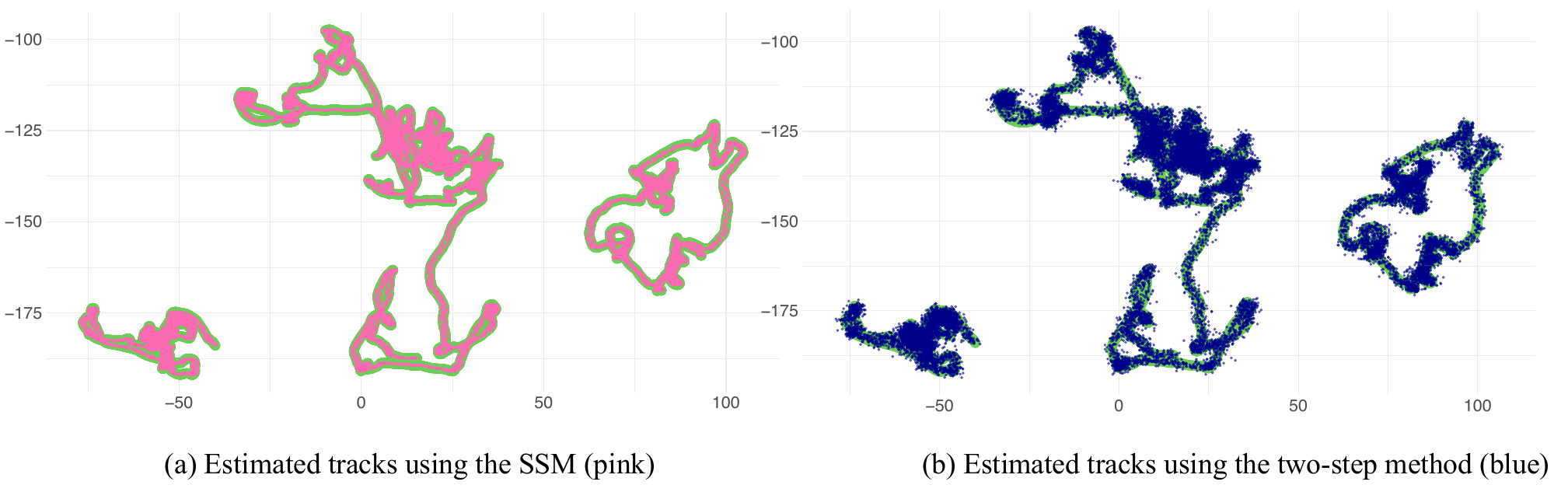} 

\end{figure}

\begin{figure}[H]
\captionsetup{font=Large, labelfont=Large}
    \caption{Comparison of Langevin SSM and two-step method for estimating $\log{(\sigma)}$ (left) and $\log{(\gamma)}$ (right)}
    \label{fig:gammasigma}
 \label{fig:combinedsigmagamma}
    \begin{subfigure}[b]{0.9\textwidth}  
        \hspace{-1cm}        \includegraphics[width=18cm, height=6.5cm]{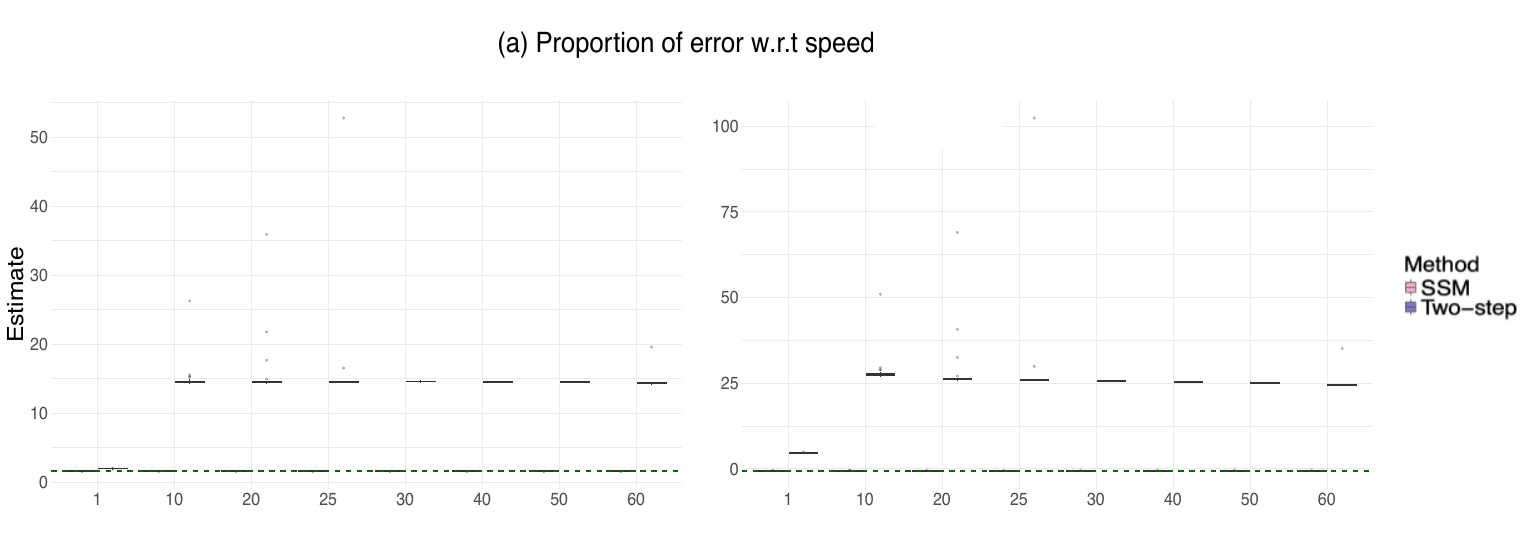}
        \caption*{}
        \label{fig:sig1}
    \end{subfigure}
     \begin{subfigure}[b]{0.9\textwidth}  
        \hspace{-1cm}
        \includegraphics[width=18cm, height=6.5cm]{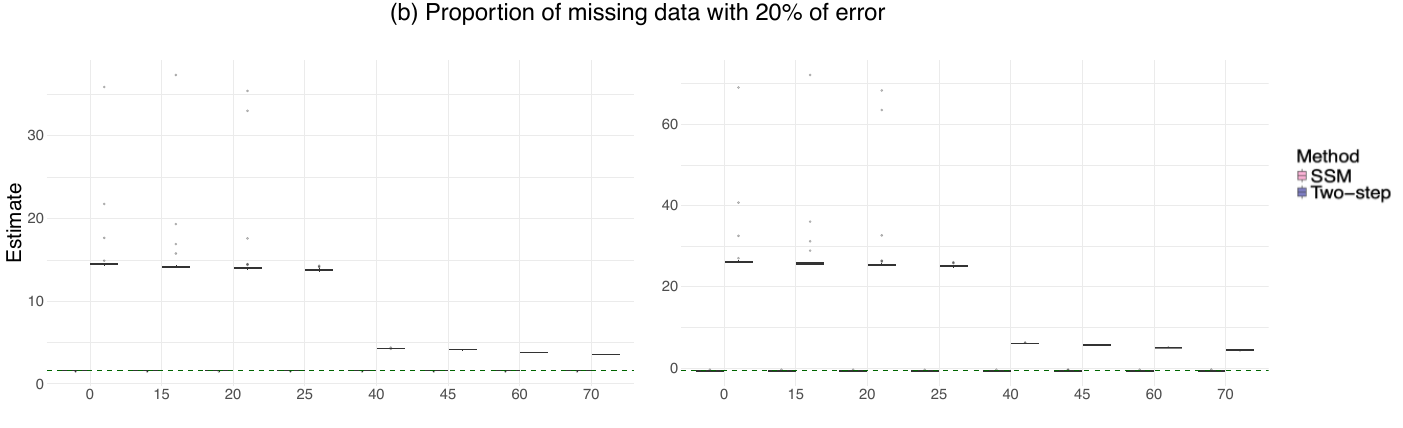}
        \caption*{}
        \label{fig:sig2}
    \end{subfigure}    
       \begin{subfigure}[b]{0.9\textwidth}  
        \hspace{-1cm}
        \includegraphics[width=18cm, height=6.5cm]{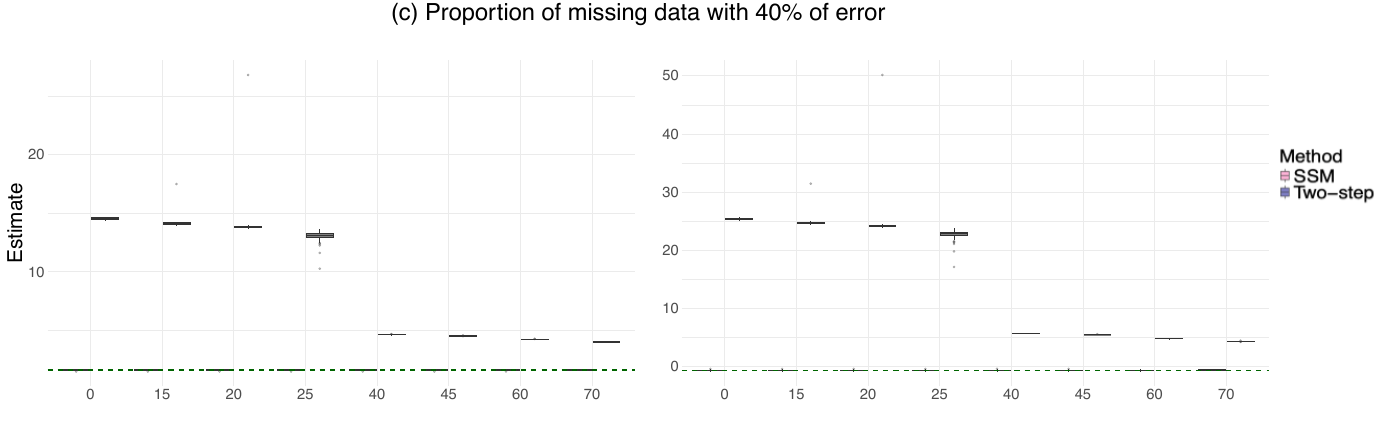}
        \caption*{}
        \label{fig:sig3}
    \end{subfigure}   
\end{figure}

\newpage

\begin{table}[H]
\captionsetup{width=19cm}
\caption[Bias, standard deviation, and coverage per parameter across different proportions of missing data with 20\% measurement error for both methods]{Simulation results: bias, standard deviation (SD), and coverage for each parameter across different proportions of missing data with 20\% measurement error with respect to speed. Values correspond to both methods (Langevin SSM, two-step method).}
\label{tab:simulation_results2}
\begin{tabular}{l c cc cc cc}
\toprule
Parameter & Proportion of missing data & \multicolumn{2}{c}{Bias} & \multicolumn{2}{c}{SD} & \multicolumn{2}{c}{Coverage} \\
\cmidrule(lr){3-4} \cmidrule(lr){5-6} \cmidrule(lr){7-8}
 & & \textbf{SSM} & \textbf{two-step} & \textbf{SSM} & \textbf{two-step} & \textbf{SSM} & \textbf{two-step} \\
\midrule
$\beta_1$ & 15 & 0.08 & 2.18 & 0.37 & 0.56 & 0.96 & 0.00 \\
 & 20 & 0.10 & 2.21 & 0.36 & 0.56 & 0.95 & 0.00 \\
 & 25 & 0.10 & 2.25 & 0.37 & 0.53 & 0.92 & 0.00 \\
 & 30 & 0.10 & 2.21 & 0.36 & 0.57 & 0.98 & 0.00 \\
 & 40 & 0.12 & 2.18 & 0.37 & 0.58 & 0.88 & 0.02 \\
 & 45 & 0.13 & 2.19 & 0.37 & 0.59 & 0.86 & 0.00\\   
 & 60 & 0.25 & 2.25 & 0.37 & 0.56 & 0.85 & 0.009 \\ 
 & 70 & 0.18 & 2.35 & 0.37 & 0.53 & 0.82 & 0.00 \\
\midrule
$\beta_2$ & 15 & -0.10 & -3.24 & 0.53 & 0.55 & 0.92 & 0.00 \\
 & 20 & -0.11 & -3.25 & 0.52 & 0.55 & 0.95 & 0.00 \\
 & 25 & -0.11 & -3.29 & 0.53 & 0.53 & 0.94 & 0.00 \\
 & 30 & -0.13 & -3.27 & 0.52 & 0.54 & 0.95 & 0.00 \\
 & 40 & -0.16 & -3.24 & 0.52 & 0.54 & 0.94 & 0.00 \\
 & 45 & -0.16 & -3.24 & 0.53 & 0.55 & 0.94 & 0.00 \\
  & 60 & -0.21 & -3.33 & 0.52 & 0.51 & 0.92 & 0.00\\ 
   & 70 & -0.25 & -3.43 & 0.53 & 0.53 & 0.89 & 0.00 \\
\midrule
$\beta_3$ & 15 & -0.07 & -2.73 & 0.44 & 0.44 & 0.97 & 0.00 \\
 & 20 & -0.08 & -2.74 & 0.43 & 0.44 & 0.97 & 0.00 \\
 & 25 & -0.08 & -2.78 & 0.44 & 0.43 & 0.97 & 0.00 \\
 & 30 & -0.09 & -2.76 & 0.43 & 0.46 & 0.97 & 0.00 \\
 & 40 & -0.12 & -2.75 & 0.42 & 0.45 & 0.96 & 0.00 \\
 & 45 & -0.11 & -2.76 & 0.44 & 0.44 & 0.95 & 0.00 \\
 & 60 & -0.14 & -2.79 & 0.42 & 0.49 & 0.95 & 0.00 \\
 & 70 & -0.20 & -2.95 & 0.43 & 0.48 & 0.94 & 0.00 \\

\bottomrule
\end{tabular}
\end{table}

\newpage

\begin{table}[H]
\captionsetup{width=19cm}
\caption[Bias, standard deviation, and coverage per parameter across different proportions of missing data with 40\% measurement error for both methods]{Simulation results: bias, standard deviation (SD), and coverage for each parameter across different proportions of missing data with 40\% measurement error with respect to speed. Values correspond to both methods (Langevin SSM, two-step method).}
\label{tab:simulation_results3}
\begin{tabular}{l c cc cc cc}
\toprule
Parameter & Proportion of missing data & \multicolumn{2}{c}{Bias} & \multicolumn{2}{c}{SD} & \multicolumn{2}{c}{Coverage} \\
\cmidrule(lr){3-4} \cmidrule(lr){5-6} \cmidrule(lr){7-8}
 & & \textbf{SSM} & \textbf{two-step} & \textbf{SSM} & \textbf{two-step} & \textbf{SSM} & \textbf{two-step} \\
\midrule
$\beta_1$ & 15 & 0.12 & 2.69 & 0.38 & 0.38 & 0.95 & 0.00 \\
 & 20 & 0.14 & 2.72 & 0.38 & 0.38 & 0.92 & 0.00 \\
 & 25 & 0.14 & 2.76 & 0.38 & 0.40 & 0.91 & 0.00 \\
 & 30 & 0.13 & 2.72 & 0.38 & 0.39 & 0.92 & 0.00 \\
 & 40 & 0.18 & 2.70 & 0.38 & 0.40 & 0.89 & 0.00 \\
 & 45 & 0.19 & 2.64 & 0.38 & 0.43 & 0.90 & 0.00 \\
  & 60 & 0.25 & 2.76 & 0.38 & 0.37 & 0.90 & 0.00 \\
   & 70 & 0.29 & 2.83 & 0.38 & 0.34& 0.87 & 0.00 \\
\midrule
$\beta_2$& 15 & -0.16 & -3.99 & 0.52 & 0.37 & 0.90 & 0.00 \\
 & 20 & -0.17 & -4.05 & 0.51 & 0.37 & 0.91 & 0.00 \\
 & 25 & -0.17 & -4.04 & 0.51 & 0.38 & 0.90 & 0.00 \\
 & 30 & -0.19 & -4.05 & 0.51 & 0.39 & 0.92 & 0.00 \\
 & 40 & -0.25 & -3.93 & 0.52 & 0.38 & 0.89 & 0.00 \\
 & 45 & -0.25 & -3.93 & 0.48 & 0.37 & 0.88 & 0.00 \\
 & 60 & -0.35 & -4.05 & 0.51 & 0.37 & 0.86 & 0.00 \\
 & 70 & -0.42 & -4.18 & 0.53 & 0.33 & 0.81 & 0.00 \\
\midrule
$\beta_3$& 15 & -0.12 & -3.37 & 0.44 & 0.33 & 0.93 & 0.00 \\
 & 20 & -0.14 & -3.40 & 0.44 & 0.34 & 0.92 & 0.00 \\
 & 25 & -0.14 & -3.39 & 0.44 & 0.34 & 0.92 & 0.00 \\
 & 30 & -0.15 & -3.41 & 0.44 & 0.32 & 0.94 & 0.00 \\
 & 40 & -0.20 & -3.30 & 0.42 & 0.34 & 0.93 & 0.00 \\
 & 45 & -0.18 & -3.34 & 0.44 & 0.36 & 0.93 & 0.00 \\
& 60 & -0.26 & -3.44 & 0.42 & 0.37 & 0.90 & 0.00\\
 & 70 & -0.33 & -3.56 & 0.44 & 0.34 & 0.84 & 0.00 \\

\bottomrule
\end{tabular}
\end{table}

\newpage

\begin{table}[H]
\captionsetup{width=19cm}
\caption[Bias, standard deviation, and coverage per parameter across different percentages of measurement for both methods]{Simulation results: bias, standard deviation (SD), and coverage for each parameter across different percentages of measurement error with respect to speed. Values correspond to both methods (Langevin SSM, two-step method).}
\label{tab:simulation_results4}
\begin{tabular}{l c cc cc cc}
\toprule
Parameter & Percentage of error w.r.t speed & \multicolumn{2}{c}{Bias} & \multicolumn{2}{c}{SD} & \multicolumn{2}{c}{Coverage} \\
\cmidrule(lr){3-4} \cmidrule(lr){5-6} \cmidrule(lr){7-8}
 & & \textbf{SSM} & \textbf{two-step} & \textbf{SSM} & \textbf{two-step} & \textbf{SSM} & \textbf{two-step} \\
\midrule
$\beta_1$ & 1 & 0.03 & -0.29 & 0.35 & 1.21 & 0.95 & 0.99 \\
 & 10 & 0.07 & 1.69 & 0.36 & 0.84 & 0.96 & 0.21 \\
 & 20 & 0.08 & 2.19 & 0.47 & 0.55 & 0.96 & 0 \\
 & 25 & 0.09 & 2.36 & 0.37 & 0.49 & 0.95 & 0 \\
 & 30 & 0.11 & 2.55 & 0.38 & 0.42 & 0.95 & 0 \\
 & 40 & 0.12 & 2.69 & 0.38 & 0.38 & 0.85 & 0 \\
 & 50 & 0.14 & 2.84 & 0.39 & 0.37 & 0.90 & 0 \\
& 60 & 0.17 & 2.96 & 0.39 & 0.35 & 0.90 & 0 \\

\midrule
$\beta_2$ & 1 & -0.02 & -0.54 & 0.52 & 1.19 & 0.97 & 0.98 \\
 & 10 & -0.07 & -2.35 & 0.53 & 0.82 & 0.97 & 0.08 \\
 & 20 & -0.10 & -3.24 & 0.53 & 0.55 & 0.92 & 0 \\
 & 25 & -0.11 & -3.51 & 0.54 & 0.50 & 0.92 & 0 \\
 & 30 & -0.14 & -3.79 & 0.52 & 0.44 & 0.91 & 0 \\
 & 40 & -0.16 & -3.99 & 0.50 & 0.37 & 0.90 & 0 \\
 & 50 & -0.20 & -4.21 & 0.50 & 0.34 & 0.90 & 0 \\
 & 60 & -0.26 & -4.37 & 0.50 & 0.31 & 0.88 & 0 \\
\midrule
$\beta_3$ & 1& -0.01 & -0.28 & 0.42 & 1.18 & 0.97 & 0.99 \\
 & 10 & -0.05 & -2.04 & 0.43 & 0.66 & 0.97 & 0.09 \\
 & 20 & -0.07 & -2.73 & 0.44 & 0.44 & 0.97 & 0 \\
 & 25 & -0.07 & -2.95 & 0.44 & 0.44 & 0.97 & 0 \\
 & 30 & -0.10 & -3.22 & 0.45 & 0.37 & 0.94 & 0 \\
 & 40 & -0.12 & -3.37 & 0.44 & 0.33 & 0.93 & 0 \\
 & 50 & -0.15 & -3.53 & 0.44 & 0.30 & 0.93 & 0 \\
 & 60 & -0.20 & 2.66 & 0.45 & 0.26 & 0.90 & 0 \\
\bottomrule
\end{tabular}
\end{table}
\subsection*{Case study}
\subsubsection*{Relationship to the \cite{delporte_spatial_2025} model}
The method of \cite{delporte_spatial_2025}, which enforces spatial constraints within the SDE of the Langevin movement model, is closely related to ours. Specifically, our inclusion of $d_{\text{water}}^{2}$ as a covariate in $\pi(\boldsymbol{\mu})$ approximates their exact formulation, with $\beta_{\text{water}}=1/{2\lambda}$ where $\lambda$ is their penalty term. Differences in performance between the two approaches are likely attributable to different modeling choices. In particular, \cite{delporte_spatial_2025} use a potential surface with attraction towards activity centers, which further pulls locations towards the center of the feasible domain and may help keep them away from land. However, such a hotspot may not be ecologically appropriate for narwhal in Qikiqtaaluk. Additionally, their simulation study is conducted with known model parameters, thereby removing an additional source of variability. The rotational component in their SDE, however, is a particularly promising extension for species moving through spatially constrained environments such as narwhal in narrow fjords.

\subsubsection*{Additional figures}
This section provides additional figures from the case study. The Langevin SSM outperforms the two-step approach at pushing locations to water.

\newpage
\begin{figure}[H]
\captionsetup{labelfont=Large}

\centering

\noindent
\begin{minipage}[t]{\textwidth}
\caption{\Large
Raw (orange) and estimated (green) locations after fitting an Langevin SSM.
}
 
\label{fig:langevin_locations_panels}
\hspace{-5.5cm}
  \begin{minipage}[t]{0.5\textwidth}
   \caption*{\hspace{3.5cm}\text{(a) locations remaining on land}
    \text{\hspace{3.5cm}(raw in orange, estimated in green)}
    \text{\hspace{3.5cm}with associated uncertainty ellipses}}
   \vspace{-0.4cm}
\hspace{3cm}
    \includegraphics[width=7.3cm,height=7.85cm]{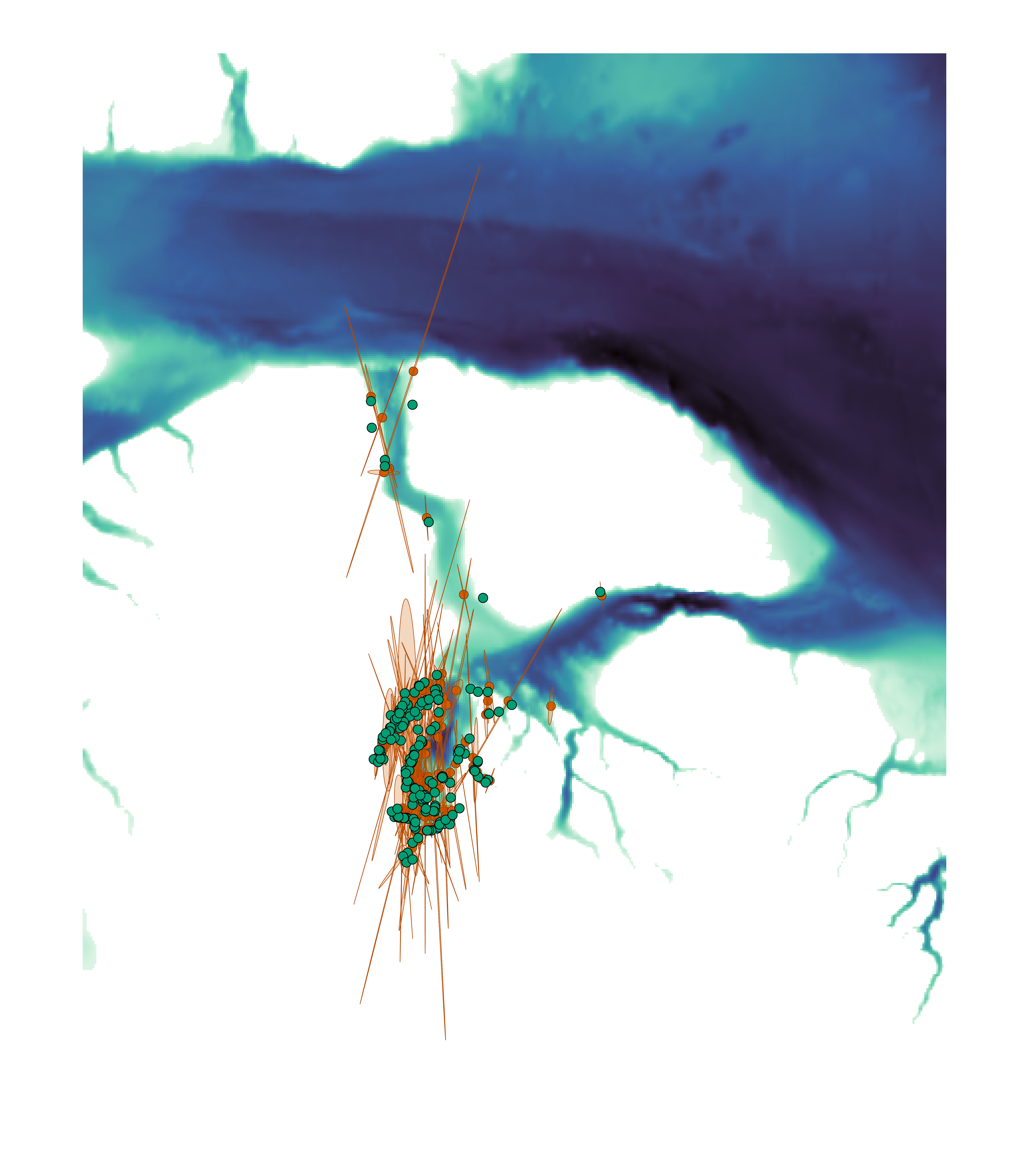}
     \end{minipage}
  \hfill
  \begin{minipage}[t]{0.4\textwidth}
    \caption*{\hspace{2cm}\text{(b) locations originally on water (orange)}
    \text{\hspace{2cm} that were pushed onto land (green)}}
    \hspace{1.5cm}
    \includegraphics[width=7.3cm,height=7.85cm]{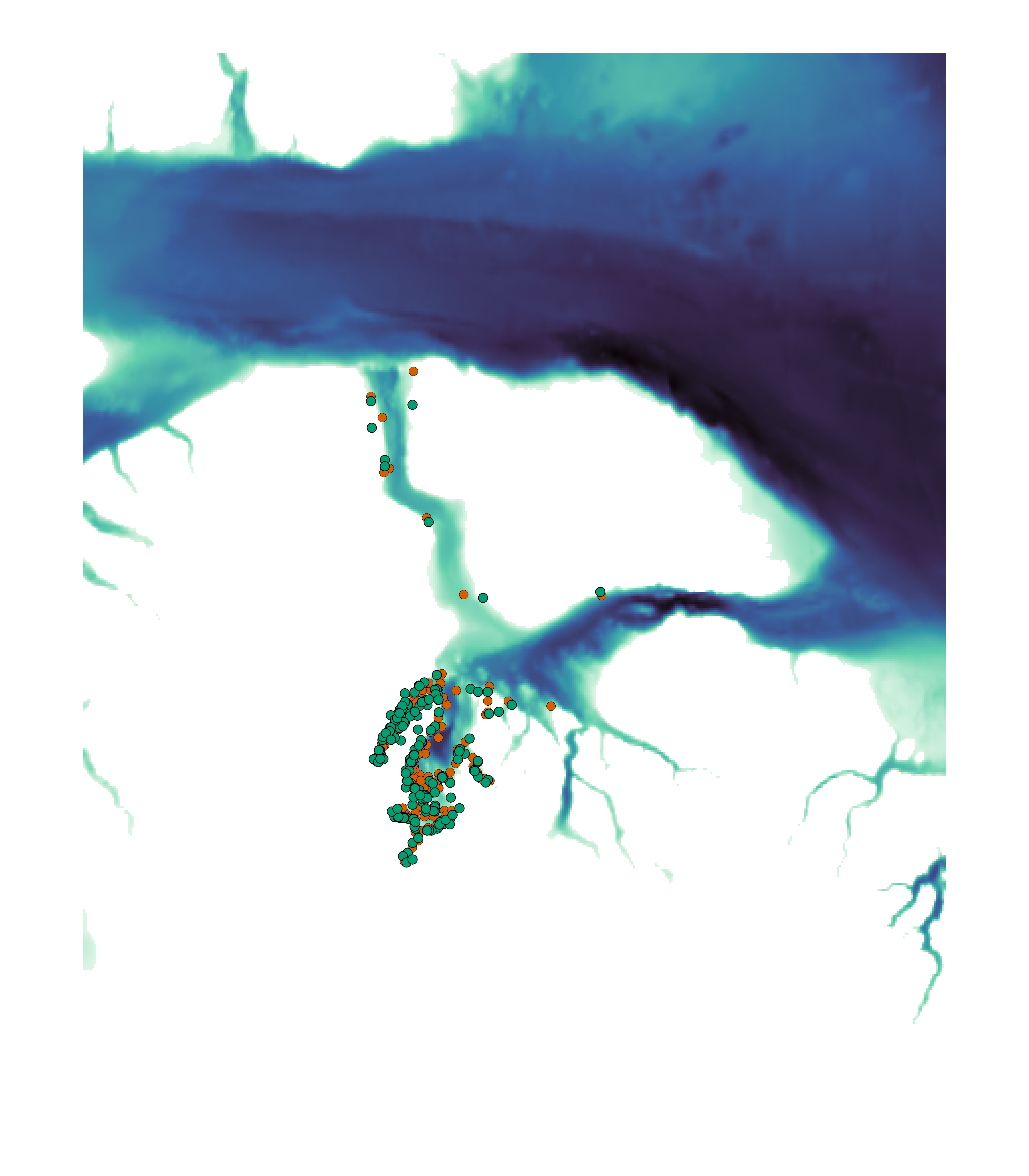}
    \end{minipage}
  \hfill
  \begin{minipage}[t]{0.39\textwidth}
    \caption*{ \hspace{2.5cm}\text{(c) locations originally on land (orange)}
    \text{\hspace{2.5cm}that were pushed onto water (green)}}
    \vspace{0.2cm}
    \hspace{1.5cm}
    \includegraphics[width=7.3cm,height=7.6cm]{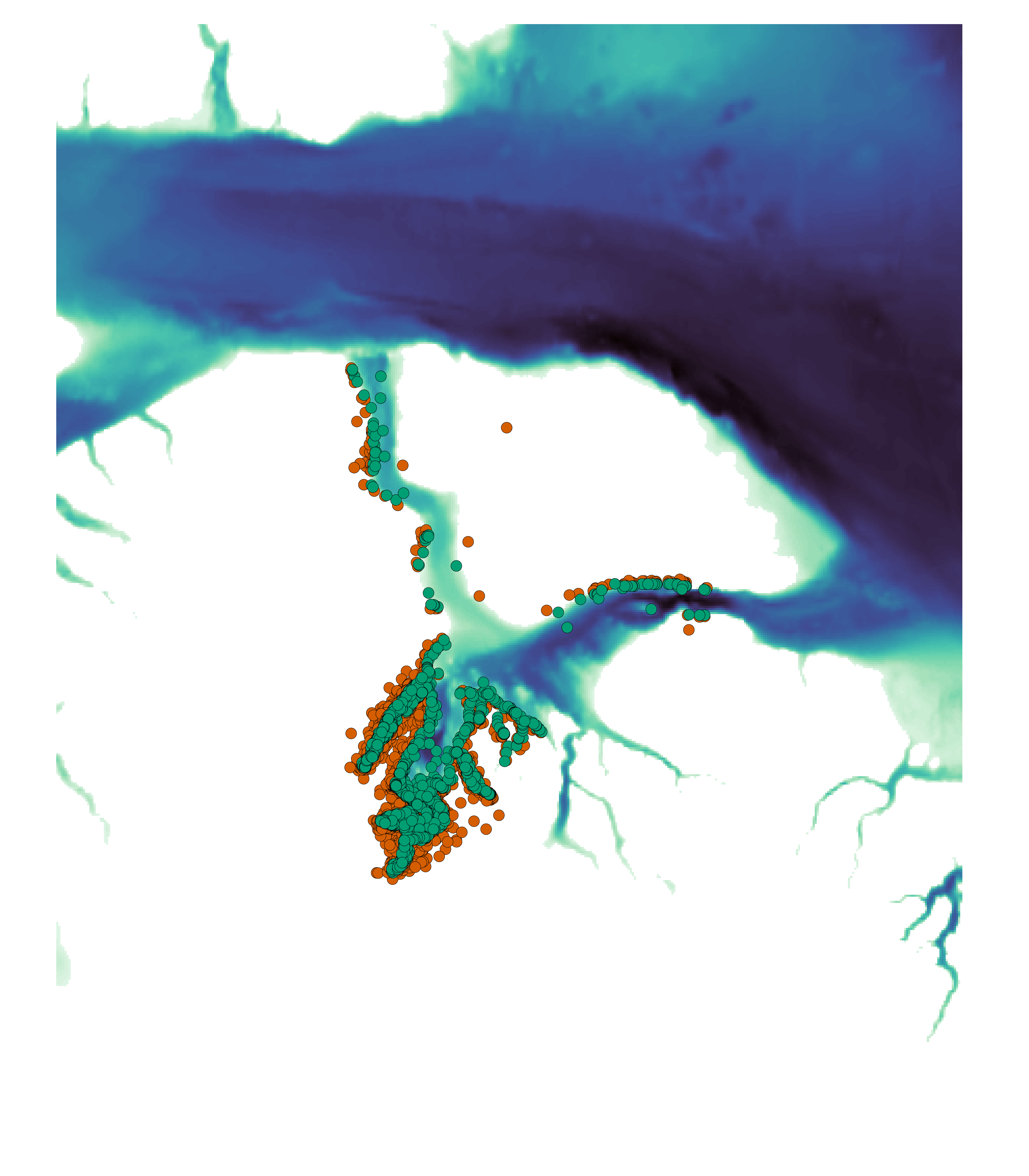}
    \end{minipage}
\end{minipage}

\end{figure}
The locations moved from water to land are largely concentrated in areas of complex coastal geometry, such as fjords, where the land-water boundary is intricate and small displacements can cross it. When the error ellipses do not overlap with any water, the penalty cannot push the associated locations toward water (Figure \ref{fig:langevin_locations_panels}a). The uncertainty ellipses shown in Figure \ref{fig:langevin_locations_panels}a correspond to the original Argos and Fastloc GPS estimates and were not adjusted using the estimated correction factors. Since $\hat{\psi}\approx5$ and $(\hat{\tau}_1,\hat{\tau}_2) = (3.9,4.1)$, the Langevin SSM effectively assumed larger uncertainty ellipses for the observation equation.

In addition, the squared distance-to-water penalty produces a steep gradient for locations far from the shore, but this gradient becomes nearly null close to complex shorelines, such as within narrow fjords. As a result, some positions may be insufficiently corrected. This mechanism also likely explains why certain locations that were originally on water but very close to land were pushed onto land, as the penalty in these areas was not strong enough to counteract spurious movement toward land. While a linear distance penalty could improve adjustments near the coast, its gradient is too weak to correct distant terrestrial points, often pushing them beyond the study area rather than into viable aquatic habitat. 

\subsection{Additional simulations}
\label{app:additional simulations}
\subsubsection*{Spatial constraint in the Langevin SSM}
\begin{figure}[H]
\hspace{-2.5cm}
\includegraphics[width=20cm,height=15cm]{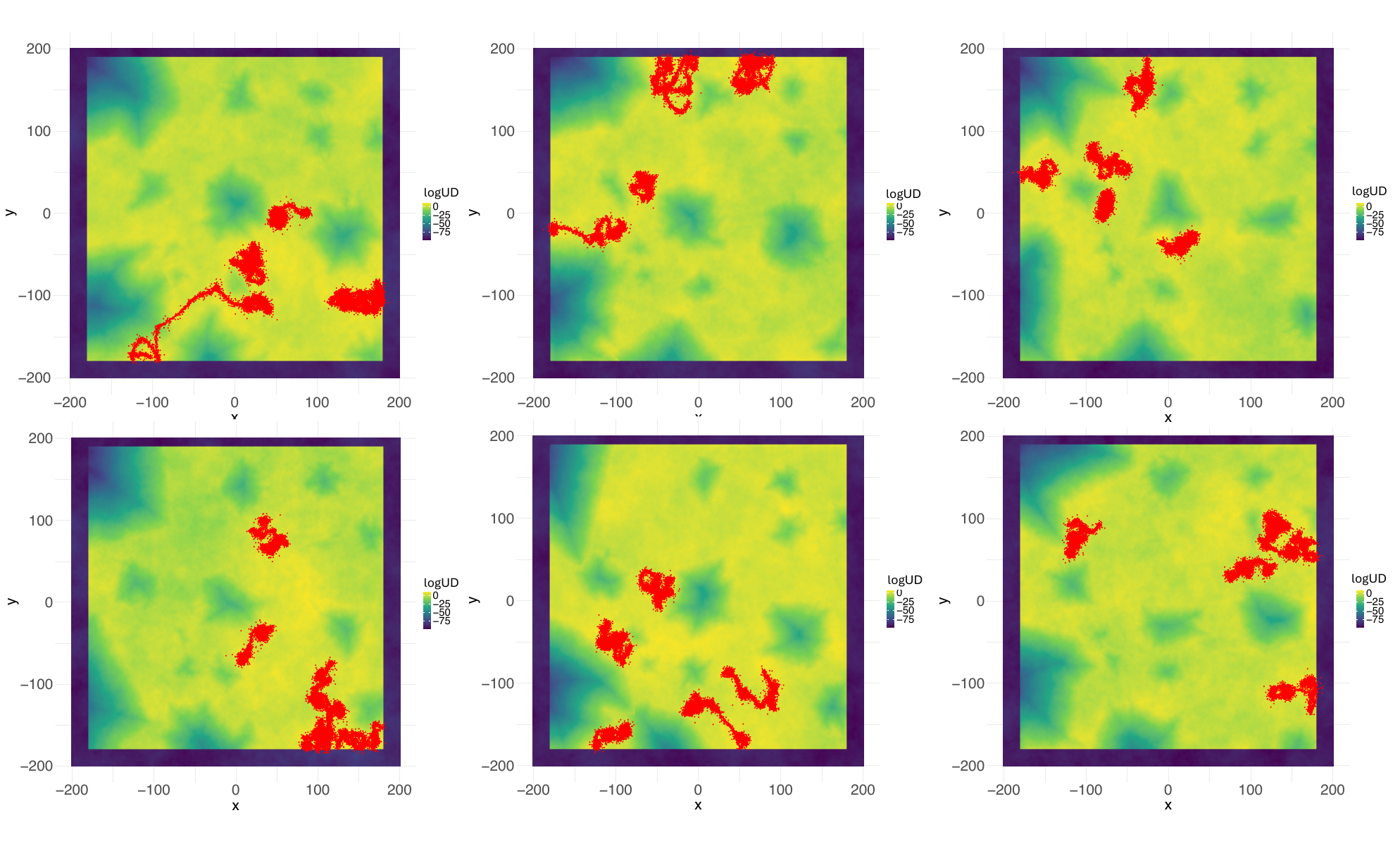}
      \caption{Six simulated tracks with $60\%$ measurement error with respect to speed}
    \label{simulationpenaltyplot}

\end{figure}

We conducted a simulation study to assess whether habitat selection patterns are preserved and accurately recovered when a spatial constraint is incorporated into the model as in Equation \eqref{eq:penalized_likelihood}. We use the same 
framework as in the main simulation study, with an additional covariate consisting of fourteen polygons of varying sizes representing landmasses, together with the boundary of the study area buffered by 20 units to represent the surrounding coastline, to which a strongly negative selection coefficient ($-100$) was assigned to enforce strict avoidance. Movement was then simulated under high measurement error ($50\%$ and $60\%$) to ensure that a proportion of observed locations fell within the constrained regions. During estimation, a negative selection coefficient was included for the constraint covariate in $\pi$ in addition to the penalty written as a function of $d^2_{\text{water}}$ in Equation \eqref{eq:penalized_likelihood},  with $\psi$ fixed at its true value. We additionally examined the effect of placing a prior on $\sigma$. However, the UD was computed without the constraint covariate, as it represents a hard spatial constraint rather than a genuine avoidance process, yielding the same UD as in the previous simulation study. Figure~\ref{simulationpenaltyplot} illustrates examples of the constrained UD and simulated tracks under $60\%$ measurement error (note that the figure displays the UD and observed locations only), where the dark patches represent constrained regions. On average, the proportion of land-based data in the $100$ simulated datasets is $37\%$, compared to $10\%$ in the narwhal dataset.
\newpage

\newpage
\begin{figure}[H]
        \caption{Habitat selection parameter estimates for the Langevin SSM with 
    added penalty for varying levels of measurement error ($50\%$, $60\%$), 
    with (pink) and without (blue) an additional penalty on the speed parameter ($\psi$ is fixed).}
\includegraphics[width=1\linewidth]{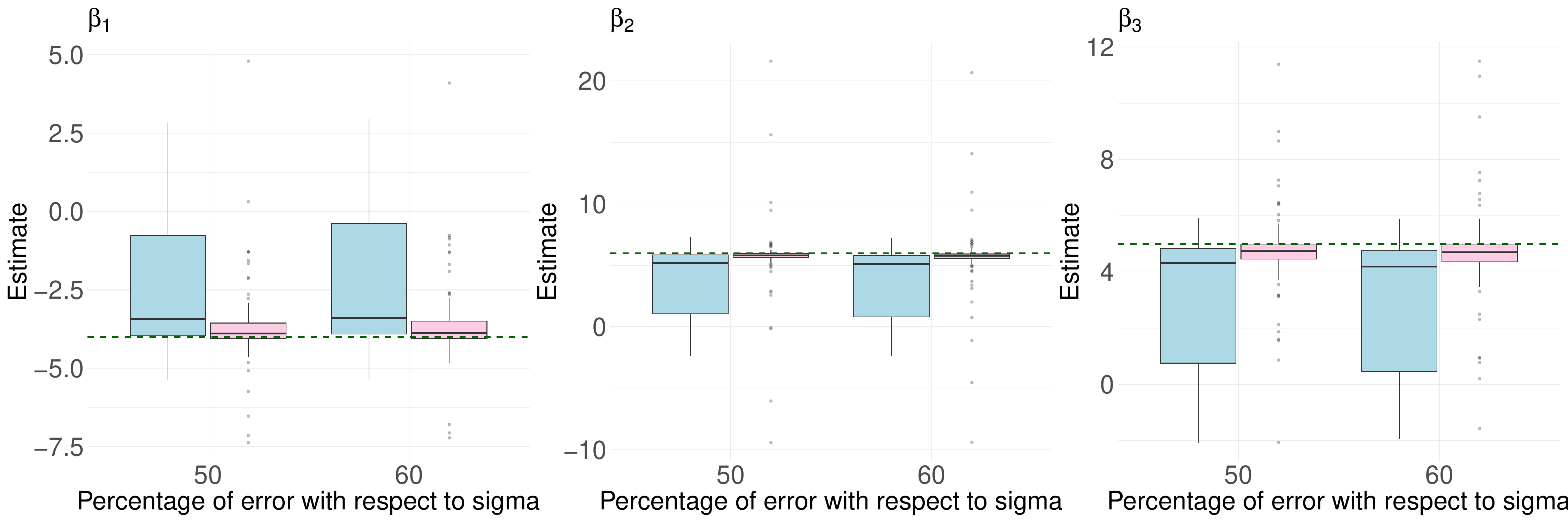}
 
    \label{fig:penalty}
\end{figure}
\newpage

\begin{figure}[H]
    \caption{Estimates of $\gamma$ and $\sigma$ for the Langevin SSM with added 
penalty for varying levels of measurement error ($50\%$ and $60\%$), without 
(blue) and with (pink) an additional penalty on the speed parameter ($\psi$ is fixed).}
    \label{fig:gammasigmaappendix}
    \begin{subfigure}{0.45\linewidth}
    \includegraphics[width=7cm,height=4.5cm]{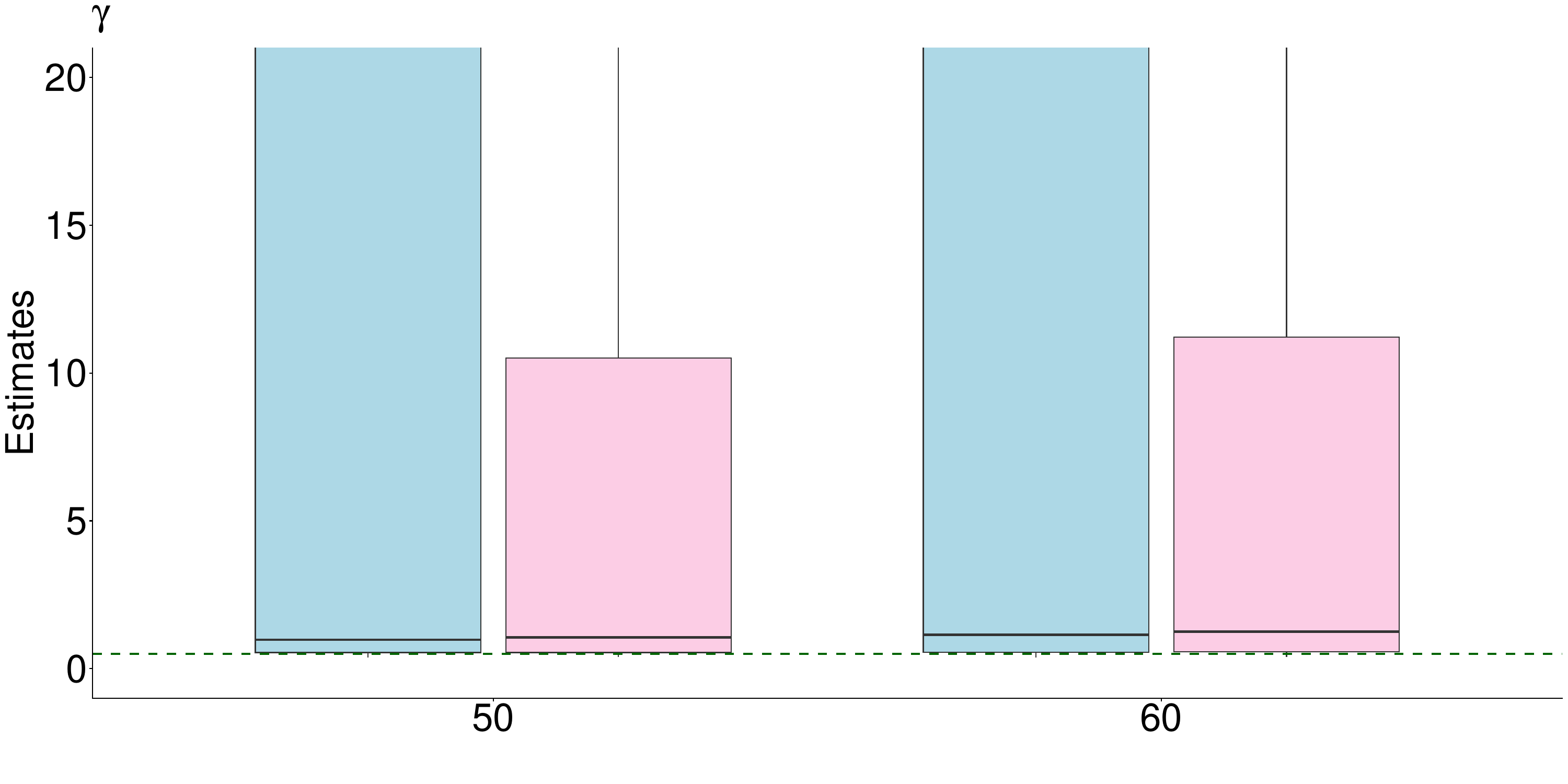}
        \caption{$\gamma$}
        \label{fig:p2_combined}
    \end{subfigure}
    \hfill
    \begin{subfigure}{0.45\linewidth}
        \includegraphics[width=7cm,height=4.5cm]{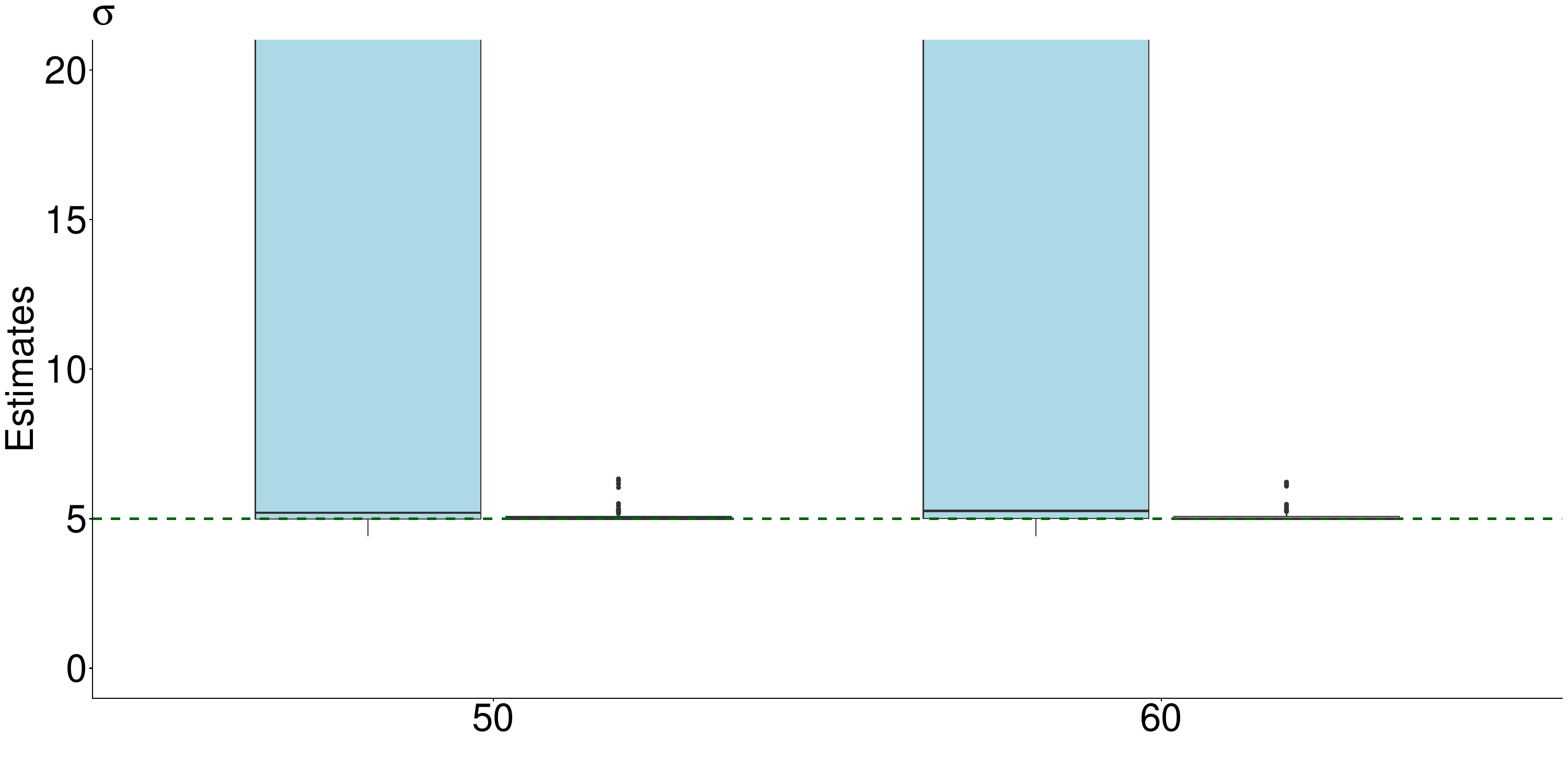}
        \caption{$\sigma$}
        \label{fig:p3_combined}
    \end{subfigure}
\end{figure}

\newpage

The introduction of a spatial constraint leads to an increase in bias and standard deviations in parameter estimates compared with the unconstrained Langevin SSM. Nevertheless, this bias remains substantially lower than that of the two-step method without spatial constraints. Under $60\%$ measurement error, the two-step approach exhibits considerably larger bias ($2.96$, $-4.37$, and $2.66$) than the constrained Langevin SSM ($1.6$, $-2.27$ and $-1.96$). An increase in standard deviation is also observed, with values of $1.91$, $2.59$, and $2.22$ for $\beta_1$, $\beta_2$, and $\beta_3$ under $50\%$ error, and $1.97$, $2.66$, and $2.29$ under $60\%$ error. The increase in variance likely reflects the additional complexity introduced by the spatial constraint: when observations fall on land, latent locations must be relocated to water, leading to greater variability in the estimation of habitat selection parameters as the model reconciles these invalid locations with the stationary distribution covariate. However, the average BA is lower for the two-step approach ($0.72$ and $0.71$) compared with the constrained Langevin SSM ($0.84$ and $0.82$). Therefore, our method is able to accommodate spatial constraints while maintaining a higher BA and lower bias than the two-step approach.

A number of simulations produced unusually large estimates of $\gamma$ and $\sigma$ ($\geq 100$, not represented in Figure \ref{fig:gammasigmaappendix} for display purposes; note that when 
one parameter explodes the other does too). We believe this is because when observations with large measurement error fall on land, the constraint forces the latent locations to be relocated towards water and, in some cases, this relocation is accommodated by inflating the movement speed, leading to larger estimates of $\sigma$ and a corresponding 
increase in $\gamma$ to preserve the habitat selection patterns (\citealp{michelot_multiscale_2024}). This phenomenon is observed in approximately $30\%$ of simulations with $50\%$ measurement 
error and $34\%$ with $60\%$ measurement error (Figure~\ref{fig:gammasigmaappendix}). This behavior also occurred in the narwhal case study, motivating our interest in addressing it. 

Adding an explicit penalty on the speed parameter improves the results. The bias in the selection parameters decreased, with values of $0.32$, $-0.32$, and $-0.27$ for $\beta_1$, $\beta_2$, and $\beta_3$, respectively, under $50\%$ measurement error and of $0.39$, $-0.36$, and $-0.30$ under $60\%$ measurement error. No simulations produced exploding ($\geq 100$) movement parameters. The decrease in bias was accompanied by an increase in average BA across simulated datasets from $0.84$ to $0.94$ with $50\%$ error and from $0.82$ to $0.94$ with $60\%$ error.

We compared the constrained Langevin SSM to the standard two-step approach without any spatial constraint, rather than to a version that removes on-land locations before fitting, as is often done in ecological studies (\citealp{auger2025including}). Some other preprocessing approaches exist, such as rerouting locations around land (e.g., \texttt{path\_rerouting} function in \texttt{aniMotum}), but we did not explore these alternatives. This reflects a key goal of our work: to avoid pre-filtering the data and retain as much information as possible. The main objective of this simulation study is to evaluate how the penalty influences the Langevin SSM estimates. Even when compared with the unconstrained two-step method, the constrained Langevin SSM consistently outperforms the two-step approach.

\subsubsection*{Two-step method with simple random walk}
\begin{figure}[H]
     \caption[Habitat selection estimates from the two-step random walk across measurement error levels]{Habitat selection parameter estimates for the two-step method with a simple random walk, with varying levels of measurement error.}
    \includegraphics[width=\linewidth]{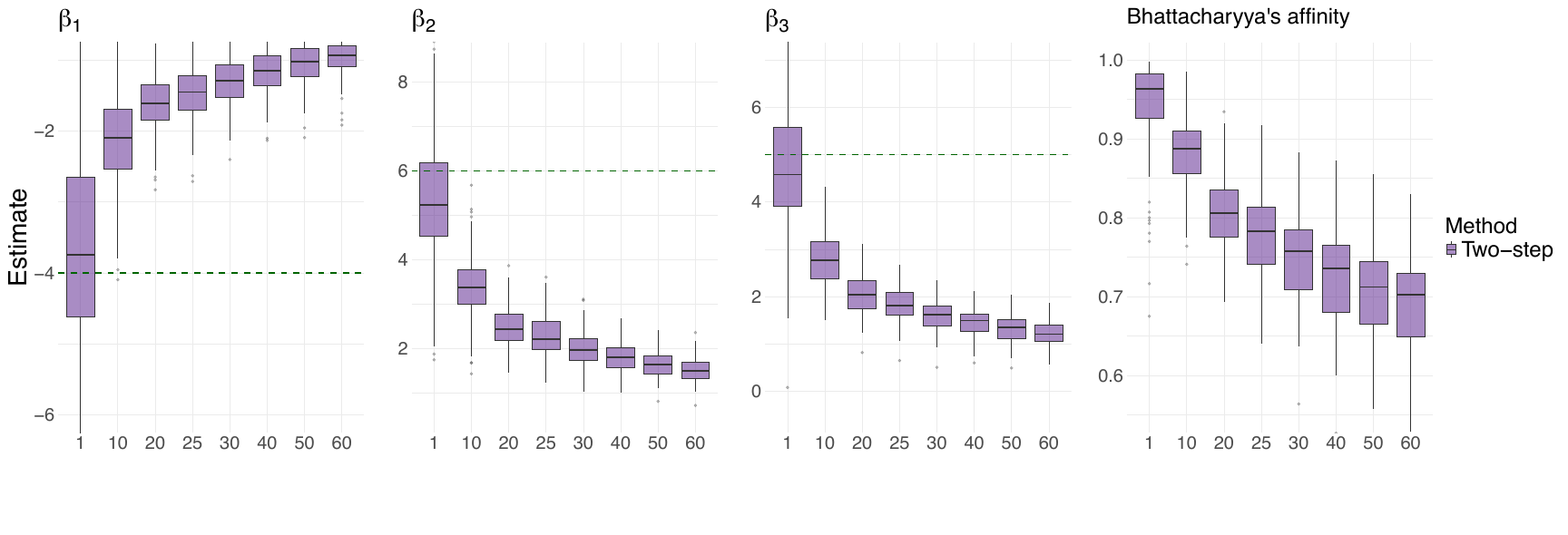}
    \label{fig:placeholder}
\end{figure}
\begin{figure}[H]
     \caption[$\sigma$ (left) and $\gamma$ (left) on the log-scale estimates from the two-step random walk across measurement error levels.]{Estimates of $\gamma$ and $\sigma$ for the two-step method with a simple random walk, with varying levels of measurement error.}
    \hspace{-2cm}
    \includegraphics[width=\linewidth]{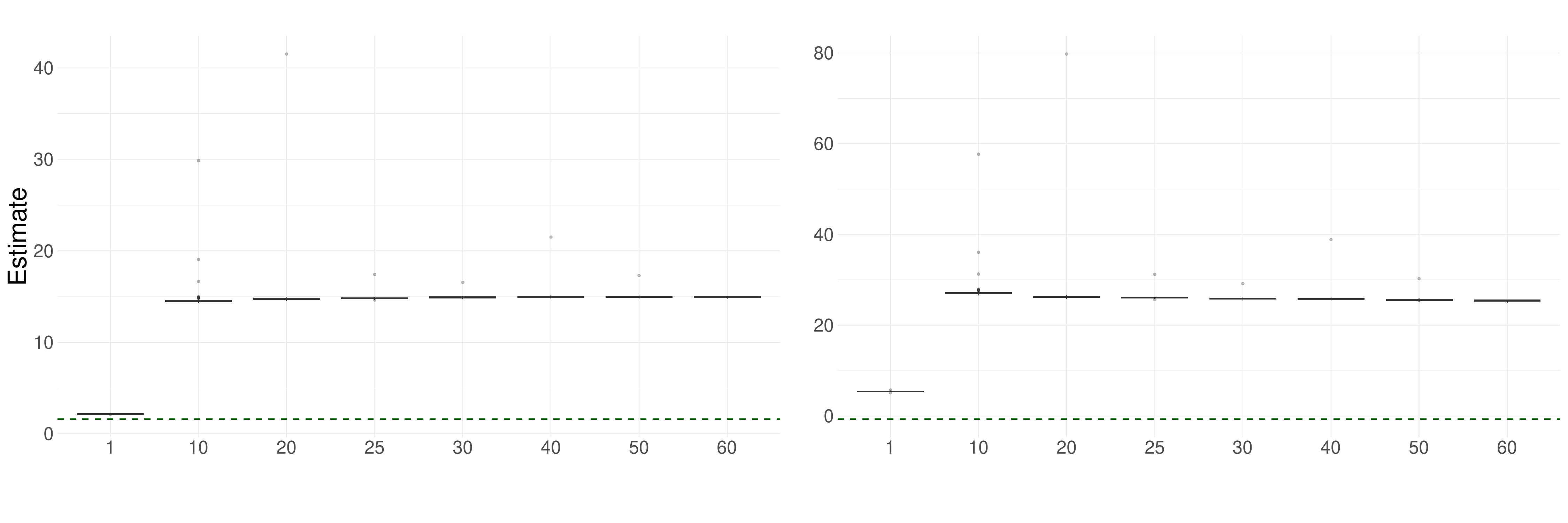}
    \label{fig:placeholder}
\end{figure}
\putbib
\end{bibunit}

\end{document}